\documentclass[11pt,preprint]{aastex}  
\newcommand{\degree}{\hbox{$^\circ$}}
\newcommand{\etal}{et\,\,al.}
\newcommand{\halpha}{H$\alpha$}
\newcommand{\hbeta}{H$\beta$}
\newcommand{\hab}{H$\alpha$/H$\beta$}
\newcommand{\gsim}
{\raise0.3ex\hbox{$>$}\kern-0.75em{\lower0.65ex\hbox{$\sim$}}}
\newcommand{\kms}{km\,sec$^{-1}$}
\newcommand{\lsim}
{\raise0.3ex\hbox{$<$}\kern-0.75em{\lower0.65ex\hbox{$\sim$}}}

\newcommand{\msun}{M$_{\odot}$}

\newcommand{\HI}{H~{\sc i}}
\newcommand{\HII}{H~{\sc ii}}
\newcommand{\HeII}{He~{\sc ii}}
\newcommand{\stpkpc}{stars\,${\cdot}$\,kpc$^{-2}$}

\newcommand{\vmi}{(V$-$I)}
\slugcomment{Scheduled to appear in AJ, December, 2003}
\begin{document}    
\title{The Recent Evolution of the Dwarf Starburst Galaxy NGC 625\\ from 
Hubble Space Telescope Imaging\footnote{Based on observations with the 
NASA/ESA Hubble Space Telescope, obtained at the Space Telescope Science 
Institute, which is operated by the Association of Universities for Research 
in Astronomy, Inc. under NASA contract No. NAS5-26555.}}
\author{John M. Cannon, Robbie C. Dohm-Palmer and Evan D. Skillman}
\affil{Department of Astronomy, University of Minnesota,\\ 116 Church St. 
S.E., Minneapolis, MN 55455}
\email{cannon@astro.umn.edu, rdpalmer@astro.umn.edu, skillman@astro.umn.edu}
\author{Dominik J. Bomans}
\affil{Astronomisches Institut, Ruhr-Universit{\" a}t Bochum,\\ 
Universit{\" a}tsstr. 150, 44780 Bochum, Germany} 
\email{bomans@astro.ruhr-uni-bochum.de}
\author{St{\' e}phanie C{\^ o}t{\' e}}
\affil{Canadian Gemini Office, Herzberg Institute of Astrophysics,\\ National 
Research Council of Canada,\\ 5071 West Saanich Road, Victoria, BC, Canada, 
V9E 2E7}
\email{stephanie.cote@nrc.ca}
\author{Bryan W. Miller}
\affil{Gemini Observatory, Casilla 603, La Serena, Chile}
\email{bmiller@gemini.edu}
\begin{abstract}

New HST/WFPC2 imaging of the dwarf starburst galaxy NGC\,625 is presented.
These data, which are 80\%\ complete to V and I magnitudes of 26.0 and 25.3
respectively, allow us to study the recent star formation history 
of NGC\,625.  Using outlying red giants stars, we derive a tip of the 
red giant branch (TRGB) distance modulus of 27.95$\pm$0.07. This 
corresponds to a distance of 3.89$\pm$0.22 Mpc, placing this system on 
the far side of the Sculptor Group.  NGC\,625 has a well-defined radial
stellar population gradient, evidenced by a central concentration of 
young main sequence stars and an RGB/AGB ratio that increases with
galactocentric distance.  The prominent AGB is very red, similar to 
the population found in the Local Group dIrr NGC\,6822.  The RGB stars
can be detected out far from the central star forming activity and 
show an elliptical distribution in agreement with the galaxy's outer \HI\
distribution.

Using \halpha\ and \hbeta\ narrow band imaging and previous optical 
spectroscopy we identify substantial and varying internal extinction 
associated with the central active star formation regions.  This 
extinction, which varies from A$_V$ $=$ 0.0 to 0.6 magnitudes,
hampers efforts to derive a detailed recent star formation history.
To better understand the effects of internal extinction on the 
analysis of young stellar populations, synthetic models are presented which,
for the first time, examine and account for this effect.  Using
the luminous blue helium burning stars, we construct a simple model of the
recent ($<$ 100 Myr) star formation in which an elevated but declining star
formation rate has been present over this entire period.  
The is at odds with the presence of spectroscopic W$-$R features in the 
major star formation region which imply a short duration ($\le$ 5 Myr)
for the recent starburst.  This suggests that starbursts displaying W$-$R 
features are not necessarily all of a short duration.  
Finally, we speculate on the possible causes of the present burst of star
formation in this apparently isolated galaxy, and compare it to other 
nearby, well-studied dwarf starburst systems.

\end{abstract}						


\section{Introduction}
\label{S1}

Nearby dwarf star-forming galaxies serve as important benchmarks 
in theories of the formation of galaxies at high redshifts, and therefore are 
vital laboratories in which the process of star formation can be studied at 
high spatial resolution ({Hodge 1989}\nocite{hodge89}, {Mateo 
1998}\nocite{mateo98}, and references therein).  Detailed studies of these 
galaxies can help to answer many questions not only about the process of star 
formation itself, but also about the process of structure formation.  By 
observing signatures of the star formation process in such galaxies, where 
large-scale galactic dynamics are absent or less important than in spiral 
galaxies, we can attempt to gauge the importance of stochastic processes in 
the evolution of low-mass star-forming galaxies.

In particular, by observing the resolved stellar populations of such objects, 
we can directly derive their recent star formation histories.  This information
allows us to constrain the nature of the star formation processes that have 
been dominant recently, including duration and strength of bursts, fiducial 
sizes and masses of star formation regions, and the temporal and spatial 
behavior of star formation throughout the galaxy.  We can then place the 
galaxies into context amongst different classes of star-forming galaxies.  For
example, very active starburst episodes might place star-forming galaxies in 
the class of blue compact dwarfs (BCD's - {Searle \& Sargent 1972}\nocite{searle72};
{Searle, Sargent, \& Bagnuolo 1973}\nocite{searle73}).  A star 
formation history that is less bursty in nature might be more representative 
of the class of dIrr galaxies.  Environmental factors likely also play an 
important role in the evolution of dwarf galaxies (see, e.g., {Skillman, 
C{\^ o}t{\' e}, \& Miller 2003a}\nocite{skillman03a} (hereafter SCM03a), and 
references therein), but these effects are not yet fully understood.  For example, 
location in a dense environment might result in a heightened present or 
recent star formation rate, or the distinctive signature of ram pressure stripping 
\citep[e.g.,][]{bureau02}.

NGC\,625 is an intriguing nearby dwarf galaxy in the Sculptor Group (or South
Polar Group), the nearest group of galaxies outside the Local Group 
({Karachentsev \etal\ 2003}\nocite{karachentsev03b}, and references therein).  
The spectroscopic studies of {Marlowe \etal\ (1997; hereafter 
MMHS)}\nocite{marlowe97} and \cite*{marlowe99} revealed strong
starburst activity, with values of star formation rate and \halpha\ equivalent 
width comparable to (but less than) those found in the prototypical local dwarf 
starbursts NGC\,1705 and NGC\,5253. The ROSAT discovery of faint x-ray 
emission \citep{bomans98} implied a large content of hot (T $\sim$ 10$^6$ K) 
gas in the halo.  Coupled with the ground-based \halpha\ imaging of 
{SCM03a}\nocite{skillman03a} and new XMM-NEWTON imaging of the galaxy {(Bomans,
Cannon, \& Skillman 2003)}\nocite{bomans03}, diffuse \halpha\ emission is 
detected away from the disk and coincident with the soft x-ray emission.  This 
leads to the interesting conclusion that this nearby galaxy is likely driving 
an outflow of hot gas from the major star formation region.  In Table~\ref{t1},
we summarize salient properties of the galaxy, as drawn from the literature.

\placetable{t1}

As this low-mass galaxy presents a low-foreground reddening sightline to a 
nearby major star formation region, it serves as an ideal galaxy in which to study
the starburst phenomenon in its own right.  We adopt the definition of 
\citet{heckman98conf}, where a starburst galaxy is any system that contains a 
spatially concentrated star formation region that dominates the overall 
luminosity of the galaxy, and that implies a gas depletion timescale much 
shorter than the age of the universe.  Starbursts are thus a brief but 
important site for massive star formation locally \citep[$\gsim$ 25\%\ of 
massive star formation occurring therein;][]{heckman98conf}, and may have been
even more important in previous epochs.  They play an integral role in the 
evolution of galaxies through their interaction with the ISM.  Furthermore, 
these galaxies represent a link between star formation and environment, as 
many are triggered by interactions, and some appear to vent metals into the 
IGM (e.g., NGC\,1569; {Martin, Kobulnicky, \& Heckman 2002}\nocite{martin02}).
Thus, NGC\,625 is a representative laboratory for the investigation of 
the details of an important mode of the star formation process. In 
\S~\ref{S9} we draw further comparisons between NGC\,625 and nearby 
starbursts.  

This paper is organized as follows.  In \S~\ref{S2} we describe the 
observations and data reduction procedures, and in \S~\ref{S3} the analysis
is presented.  The distance of the galaxy is derived from the tip of the 
red giant branch in \S~\ref{S4}.  In \S~\ref{S5} we discuss 
the models used to analyze the recovered photometry, the methods used to 
isolate the various stellar populations within NGC\,625, and our new method 
of simulating the effects of differential extinction in stellar populations.  
We address the young and old stellar populations, respectively, in 
\S\S~\ref{S6} \&\ \ref{S7}, with particular attention paid to the rigorous 
burst of star formation which this relatively isolated galaxy is undergoing, 
and the simulations used to interpret this young stellar population.  In 
\S~\ref{S8} we speculate on the triggering mechanism of the current 
starburst episode, in \S~\ref{S9} we compare NGC\,625 with other dwarf 
starburst systems, and finally in \S~\ref{S10} we draw conclusions about 
starburst galaxies in general and NGC\,625 in particular.

\section{Observations and Data Reduction}
\label{S2}

NGC\,625 was observed with the Wide Field/Planetary Camera 2 (WFPC2) aboard 
the Hubble Space Telescope (HST) on 2000, September 23 \& 24.  Images were 
obtained in four different passbands: V (F555W filter, 5200 sec), I (F814W 
filter, 10400 sec), \halpha\ (F656N filter, 800 sec) and \hbeta\ (F487N 
filter, 1200 sec).  Table~\ref{t2} summarizes the data obtained.  
Figure~\ref{figcap1} shows the HST/WFPC2 field of view superposed on a 
ground-based R-band image.  The brightest central clusters were centered on 
the PC chip to minimize stellar crowding.  The field of view 
encompasses most of the central disk and much of the eastern halo of the 
galaxy, allowing the study of stellar populations in both field and 
star-forming regions.  No dithering strategy was applied during the data 
acquisition.  The images were processed through the standard pipeline 
reductions at STScI.

\placetable{t2}

\subsection{Broad Band Images}
\label{S2.1}

The images were combined into single exposures for each passband by 
applying the anti-coincidence routine of \citet{saha96} to remove cosmic 
ray detections between pairs of images.  The success of this method, which
uses a moderate rejection threshold (4\,$\sigma$), requires
multiple pairs of images in each filter.  All images were obtained at the 
same pointing, mitigating the need for aligning successive images.  
Further reductions were not applied to the broad band images before their
analysis (e.g., removal of background levels); rather, the automated 
photometry routines applied (see \S~\ref{S3}) account for such effects.  
A four-color composite image is presented in Figure~\ref{figcap2}.
Note the high stellar density in the disk, and the major starburst
region which is displaced toward the eastern end of the disk.  Note also 
the clear presence of nebular emission and dust obscuration. 

\subsection{Narrow Band Images}
\label{S2.2}

The reduction and calibration of the narrow band images followed the 
methodology outlined in \citet{cannon02}.  We need to quantify potentially 
subtle features (e.g., variations in the \hab\ flux ratio), thus requiring 
great care that the calibration be as precise as possible.  For consistency, 
the broad band images used for the continuum subtraction (see below) were 
re-reduced via the same prescription applied to the narrow band images.  

The images were first corrected for warm pixel effects and geometric 
distortion.  Cosmic rays were removed via the IRAF\footnote{IRAF is 
distributed by the National Optical Astronomy Observatories, which are 
operated by AURA, Inc., under cooperative agreement with the National Science 
Foundation.} task CRREJ, which is reasonably effective with only two images, as
for the F656N (\halpha) and F487N (\hbeta) images.  However, some cosmic rays 
were not filtered out and these were removed manually.  An average background 
level was determined for each chip and then subtracted.  Due to the large 
extent of the galaxy, estimating the background is difficult for the broad 
band images (particularly for the PC1 and WF2).  Fortunately, due to the small
contribution of the underlying stellar continuum (see below), this error term 
will be of second order.  We discuss below the overall accuracy of our narrow 
band calibration, and its total error budget.

The broad band images were flux calibrated using standard methods.  
For the narrow band images, the system responses of the filters
at the redshifted wavelengths of the emission lines were modeled with the 
SYNPHOT package.  We adopt a heliocentric radial velocity of 405 \kms\ 
for NGC\,625, derived from the \HI\ synthesis imaging of 
{C{\^ o}t{\' e}, Carignan, \& Freeman (2000; hereafter CCF00)}\nocite{cote00}.  
The narrow band images were then corrected for differences in throughput at 
these wavelengths compared to the nominal values at the band centers.

In narrow band analysis, removing the underlying stellar continuum is 
often the most difficult step of the analysis.  \citet{dutil01} 
suggest that, regardless of quality and signal-to-noise ratio of the 
data, the precision of continuum subtraction cannot be better than 2\%\ 
in \halpha\ and 4\%\ in \hbeta.  The most effective method is suggested to be
the simple ``manual method'', and should be preferred to more sophisticated
techniques, such as $\chi^2$ minimization.  In the present analysis we apply 
a variant of this technique, which is described as follows. 

The F555W image is contaminated by both \halpha\ and \hbeta\ emission, while 
the F814W bandpass is not sensitive to either line.  Following 
\citet{cannon02} and \citet{dutil01}, the continuum subtraction was thus 
achieved recursively by scaling and subtracting the images (i.e, manually).  
When the scaled narrow band images were subtracted from the wide band images, 
we considered the continuum optimally removed when there no longer remained 
significant regions of nebular contamination in the wide band filter.  We 
attach the aforementioned 4\%\ uncertainty to this value (see \S~\ref{S3} for 
a description of the errors that affect the narrow band images). 

The continuum-subtracted narrow band images were then multiplied by the 
effective filter widths as calculated by SYNPHOT.  The final, 
continuum-subtracted emission line images are presented in 
Figures~\ref{figcap3}(a) and \ref{figcap3}(b).  Note that the \hbeta\ image, 
while a longer integration, presents the lowest signal-to-noise ratio (and 
statistically dominant error term, see \S~\ref{S3}).  Comparison to 
ground-based \halpha\ studies lends confidence to our calibration.  Our 
derived total \halpha\ flux of (3.49$\pm$0.18)$\times$10$^{-12}$ erg sec$^{-1}$
cm$^{-2}$ agrees well with that found by {SCM03a}\nocite{skillman03a} 
((3.3$\pm$0.2)$\times$10$^{-12}$ erg sec$^{-1}$ cm$^{-2}$) and by
{MMHS}\nocite{marlowe97} (3.3$\times$10$^{-12}$ erg sec$^{-1}$ cm$^{-2}$). 

\section{Analysis}
\label{S3}

\subsection{Stellar Photometry}
\label{S3.1}

The DoPHOT program \citep*{schechter93}, with modifications for the 
undersampled WFPC2 pixels \citep{saha96}, was used to derive the photometry of 
the individual stars in the V and I images.  The subtleties involved in 
applying DoPHOT to crowded WFPC2 images have been discussed at length in 
\citet{dohmpalmer97a} and \citet{dohmpalmer02a}, to which we refer the 
interested reader for specific details.  

As an inspection of Figure~\ref{figcap2} will reveal, the stellar crowding is 
high in the center of the galaxy, most noticeably in the PC and WF2 chips.  In 
an effort to curtail the detrimental effects of crowding, DoPHOT subtracts an 
analytically defined, scaled point spread function (PSF) from the image at each
star's location. Highly crowded areas will be susceptible to errors in this 
process, as subtraction residuals can be mistaken for stars.

DoPHOT's analytic PSF fitting function is controlled by four parameters: 
$\beta_4$, $\beta_6$, $\beta_8$, and the FWHM.  Following 
\citet{dohmpalmer97a}, these parameters are fitted to the brightest, most 
isolated stars in each chip.  The PSF is determined by minimizing the residuals
of subtractions of isolated stars from the images.  The best results were found
by allowing small variations in FWHM from chip to chip, but using the same 
values for $\beta_4$, $\beta_6$ and $\beta_8$ as derived in 
\citet{dohmpalmer02a}.  These shape parameters are given explicitly in 
Table~\ref{t3}.  

Aperture corrections were next performed on the DoPHOT outputs.  
Corrections were measured in an aperture of 0.5\arcsec\ for stars with small 
internal DoPHOT errors ($\leq$ 0.04 mag in I, $\leq$ 0.08 mag in V).  The
high crowding levels in some chips made it difficult to isolate numerous 
stars of such quality, and in some cases, the aperture correction for a 
chip had to be based on a relatively small number of stars.  A polynomial
function (quadratic in position) was then fit to these derived values.  
Table~\ref{t3} lists the number of stars used to calculate the aperture 
correction per chip, and the rms residual of the fit.  As expected, the 
PC chips have the highest rms value, due to the highest levels of 
crowding found there. 

\placetable{t3}

Charge transfer efficiency (CTE) corrections were applied according to the 
prescription of {Whitmore, Heyer, \& Casertano (1999)}\nocite{whitmore99} 
(which is in general agreement with the more recent calibration by 
\citet{dolphin00b}).  Photometry lists in V and I were next matched to a 
radius of 0.5 pixels.  In total, 52,188 stars were matched between the two 
filters.  More stringent cuts in these data (based on total errors in V and I 
magnitude) will be used in the analysis that follows, reducing the number of 
stars used to derive the star formation history.  Finally, the transformation 
to the V and I magnitude system was found via the equations given in 
{Holtzman \etal\ (1995\nocite{holtzman95b}, which is in good agreement with 
{Dolphin 2000}\nocite{dolphin00b}).  The complete I vs. \vmi\ and V vs. \vmi\ 
color magnitude diagrams for all detected stars are shown in 
Figure~\ref{figcap4}.

At the high galactic latitude of $-$73\degree, the foreground reddening 
toward NGC\,625 is expected to be low.  The maps of {Schlegel, Finkbeiner,
\& Davis (1998; hereafter SFD98)}\nocite{schlegel98} find a value of E(B$-$V) 
= 0.016, which we adopt in the following analysis (see also \S~\ref{S3.3}).  
Applying the extinction coefficients from the same work yields A$_V$ = 
3.25$\cdot$E(B$-$V) = 0.05 mag, and A$_I$ = 1.95$\cdot$E(B$-$V) =  0.03 mag.  
These corrections have been applied to the calibrated photometry presented 
here, and a 10\%\ error on the A$_{\lambda}$ values propagated (although this 
error is usually negligible compared to the internal DoPHOT photometric 
errors).

Foreground stellar contamination is also expected to be minimal; the models 
of \citet{ratnatunga85} predict a contamination of 2.3 stars arcmin$^{-2}$
(brighter than 27th magnitude in V) for the nearest appropriate globular 
cluster (NGC\,4147; 21\degree\ offset).  For the usable field of view of the 
WFPC2 camera, we then expect $\sim$ 12 stars to contaminate the CMD of 
NGC\,625.  Thus, we conclude that such contamination is negligible for the 
analysis at hand.

In Figure~\ref{figcap5}, we present the total photometric errors for all 
matched stars between I and V (including errors from the foreground 
reddening correction).  The V data are roughly one magnitude deeper than 
the I data.  The larger errors and greater spread of this distribution 
toward faint magnitudes is expected.  The large number of stars which 
scatter above this trend is due to the high stellar density and crowding 
throughout some regions of the images.

To test the completeness of our photometry, artificial star tests were 
performed.  The V-band luminosity function was found to fit a power law with 
exponent 0.44.  Artificial stars were created according to the analytic 
function which was used to fit the bright, isolated stars in each image. Stars
randomly distributed with this power-law distribution in V magnitude were then
added at the 2\%\ level to 100 copies of each image.  These stars were randomly
distributed in position.  Each random distribution was applied to both the V 
and I images, with \vmi\ colors randomly assigned between $-$1 and 2.  
Poisson noise was then added to each artificial star.

DoPHOT was applied to each of the artificial images using identical parameters
to those used in the science frames.  The input star lists and the artificial 
frame photometry were then matched, and the fraction of recovered stars was 
measured as a function of magnitude.  Due to the steep stellar density gradient
present in some sections of the galaxy, the resulting completeness estimates 
are calculated as a function of stellar density.  These calculations are 
discussed in the next section.  

\subsection{Isolating Crowding Effects}
\label{S3.2}

Some sections of the the V and I images suffer from severe stellar crowding, 
which will have detrimental effects on the recovered photometry.  Furthermore,
the steep stellar gradient between the disk and the halo results in  deeper
photometry in the outer parts of the galaxy.  In order to isolate and minimize
these effects, we break up the photometry into annular regions which 
will have relatively equal amounts of stellar crowding and 
background variation.  This analysis allows us to isolate crowding effects 
with a fair degree of accuracy.

In analyzing the stellar photometry, we follow the general method adopted by 
\citet{tosi01} in their study of the starburst galaxy NGC\,1705.  Here we 
separate the CMD into sections using elliptical contours of a fixed axial 
ratio (2.8:1) guided by isophotal fits to the images.  This elliptical contour 
analysis allows us to study the distribution of stars with respect to the main
disk of the galaxy.  This will be very important, for example, in the 
determination of the distance from the TRGB (see \S~\ref{S4}) and in the study
of the spatial distribution of the various stellar populations found 
throughout the galaxy (see \S\,\S~\ref{S6} \&\ \ref{S7}).   

The contours used to divide the photometry are displayed in 
Figure~\ref{figcap6}.  The galaxy is divided into six regions.  The outer stars
in the periphery of the galaxy are included in Region\,1.  Crowding levels 
increase as one approaches the main disk of the galaxy.  Region\,6 contains 
the highest stellar densities, as well as the large \HII\ regions listed in 
Table~\ref{t4} and the dust concentrations discussed in \S~\ref{S3.3} and 
listed in Table~\ref{t5}.  The individual color magnitude diagrams for each 
section of the galaxy are shown in Figure~\ref{figcap7}.    

From an inspection of Figure~\ref{figcap7}, it is clear that these areas not 
only isolate similar crowding levels, but also similar age progressions of the
stellar populations contained within.  The blue plume, which consists of main 
sequence (MS) and blue helium burning stars (BHeB), is prominent in Regions\,5
and 6, and decreases in strength as one moves away from the main disk (toward
Regions\,2 and 1).  The very red asymptotic giant branch (AGB) is evident in 
all regions, but most easily identifiable in the regions further from the disk 
(Regions\,1 and 2, for example).  The implications of such a spatial 
distribution are discussed in more detail in \S\S~\ref{S6} \&\ \ref{S7}.  The 
red giant branch (RGB) is most easily isolated in the outer regions (see 
\S~\ref{S4}), and becomes confused with the faint end of the red supergiant 
population in regions closer to the main disk of the galaxy.  

In order to compare the stellar distributions in each of the annuli, we must 
consider the relative completeness of the photometry in each region.  
Figure~\ref{figcap7} shows
that the inner regions, where stellar crowding is most severe, suffer from 
relatively high incompleteness.  
By separating our artificial star simulations into the same regions 
used to analyze the photometry, we determine that Region\,6
becomes incomplete at 0.5 (V) and 1.2 (I) magnitudes shallower than Region\,1.
In Figure~\ref{figcap7} we include the results of the completeness tests for 
each of the six regions (also discussed further in \S\S~\ref{S6} \&\ \ref{S7}).

\citet{aparicio96} emphasize that detailed crowding effects must be included 
in the distribution of artificial stars which is input into the images.  By 
separating the analysis of our observations into annuli of similar stellar 
densities (i.e., separate photometric errors and completeness measurements) we
can accomplish the same goal.  Since  our analysis will concentrate on stars 
that are primarily above the completeness limits of the photometry (e.g., red
giants, and younger stars in the blue plume) completeness estimates are not 
critical to most of our interpretation.

\subsection{Narrow Band Photometry}
\label{S3.3} 

The narrow band images were analyzed with two goals in mind.  First, we can
study the distribution of \halpha\ emission in comparison with the distribution 
of young stars.  Second, in tandem with the \hbeta\ image, we can map the 
large-scale effects of reddening throughout the disk.  This will help us to 
correct the CMD analysis of the affected populations.  

Due to the intrinsically lower background levels, the effects of charge 
transfer efficiency must be carefully considered in the analysis of HST 
narrow band imaging.  However, because of the small area of the CCD covered by
the nebular emission, the prescriptions of \citet{whitmore99} and 
\citet{dolphin00b} both give correction terms of order 1\%\ over the extent of 
the features in both \halpha\ and \hbeta, as seen in Figure~\ref{figcap3}.  
The effects of these corrections are negligible compared to the 
errors inherent in the subtraction of the continuum (see below), so 
pixel-by-pixel CTE corrections were not applied to the narrow band images 
considered here.  As previously mentioned, the foreground reddening toward 
NGC\,625 is low, E(B$-$V) = 0.016 mag.  This foreground correction has been 
applied to the narrow band images.   

{SCM03a}\nocite{skillman03a}, using ground-based images, catalog 23 \HII\ 
regions in NGC\,625. Because of the higher spatial resolution (and hence lower
surface brightness sensitivity) of the WFPC2 camera, only the four brightest 
\HII\ regions are detected (our regions A, B, C, D correspond (roughly) to 
their \HII\ regions 5, 9, 18, 4, respectively).  The properties of these four 
\HII\ regions are listed in Table~\ref{t4}. The four \HII\ regions account for
roughly  95\%\ of the total \halpha\ flux recovered from the entire galaxy.  
We note that these four \HII\ regions also account for over 90\%\ of the total
\halpha\ flux found in the ground based study.   This agreement suggests that 
the lower surface brightness \HII\ regions found in ground-based images have 
not been recovered here, and that the nature of the unrecovered \halpha\ is 
mostly discrete, low surface brightness \HII\ regions, rather than a diffuse 
background.  While we cannot completely rule out a diffuse component (perhaps 
as large as 5 - 10\%\ within our errors), this fraction is small in comparison
with the larger diffuse component (10 - 30\%) inferred for some gas rich dwarf
irregulars \citep[e.g.,][]{vanzee00a,kennicutt01}. 

\placetable{t4}

We then analyze the continuum-subtracted \halpha\ and \hbeta\ images via 
aperture photometry to address the potential contribution from internal (and 
possibly highly variant) extinction.   Apertures are used to attain emission 
line fluxes in areas of high equivalent width in both the \halpha\ and \hbeta\ 
images.  From inspection of Figure~\ref{figcap3}(b), it is clear that the low 
signal to noise ratio of the \hbeta\ image will limit our ability to use this 
technique on a widespread basis throughout the galaxy.  We list in 
Table~\ref{t5} the recovered photometry for the \HII\ regions delineated in 
Figure~\ref{figcap3}(b).  Each \HII\ region shows roughly 0.5 magnitudes of 
extinction at V.  These values are consistent with the higher values found in 
the spectroscopic observations of {Skillman, C{\^ o}t{\' e}, \& Miller 
(2003b; hereafter SCM03b)}\nocite{skillman03b}, where \hab\ ratios are 
elevated above the theoretical value of 2.85 \citep{hummer87} to as high as 
3.6. However, the spectra of {SCM03b}\nocite{skillman03b} indicate that one 
of the \HII\ regions is consistent with no reddening at all.
This suggests the presence of variable extinction within the star formation 
regions.  To test for this effect, we separated the areas of high \hbeta\ 
equivalent width into sections, and performed aperture photometry on 
each smaller area individually.  We find that extinction levels vary by $\sim$ 
0.2 magnitudes over regions as small as 1\arcsec\ (18.9 pc at the distance 
derived in \S~\ref{S4}) in the central star formation complexes.  

The presence of strong underlying stellar absorption can mimic the behavior
of dust extinction.  The large (aperture-averaged) equivalent widths derived 
here (see Table~\ref{t5}), however, suggest that underlying stellar absorption
plays only a minor role in elevating our observed \hab\ ratios from the 
theoretical value, and that it is a negligible effect.  The models of 
\cite*{gonzalezdelgado99b} demonstrate that the contribution from underlying 
absorption rises with age of the ionizing cluster, and may be as large as 
$\sim$ 10 \AA\ for clusters which are hundreds of millions of years old.  
However, the luminosity and equivalent width of the nebular emission argues 
for a much younger age, with typical contributions from underlying absorption 
of order 3 \AA, representing less than a  1\%\ effect.  Thus, we conclude that 
the bulk of the elevated \hab\ ratio is a consequence of (highly variable) 
dust extinction along sightlines into the active star formation regions.  This
varying extinction will cause our stellar photometry to be spread out in 
color and magnitude; while we may measure a star with small photometric errors
in the central regions, this does not necessarily mean that we recover its 
true absolute magnitude.  We argue later (see \S~\ref{S5}) that this effect 
widens our CMD features, rendering our ability to separate different stellar 
populations less accurate than it would otherwise be in the presence of little
(or smoothly varying) extinction.  

\placetable{t5}

It has been shown by \citet{zaritsky99} and \citet{zaritsky02} for the 
Magellanic Clouds, and by \citet{calzetti01} for starburst galaxies, that the 
contributions to internal extinction are dependent on the population being 
analyzed.  For the Magellanic Clouds, extinctions are of order several tenths 
of a magnitude larger in the young populations, while variations as large as 
one magnitude are found in starburst galaxies \citep[e.g., 
NGC\,5253;][]{calzetti97}.  Since we only have two-color information, we 
cannot address this issue in detail, but our general results using the \hab\ 
ratio seem to favor such an interpretation.  Individual line-of-sight
corrections for this effect are not 
applied; however, we note the potential uncertainty introduced as a result.  
The effects of differential extinction could be widespread throughout the disk
and not necessarily confined to the regions where we have measured them. This 
is discussed further in \S~\ref{S5}. Furthermore, we note that this correction
applies in the direction of increasing the luminosities of the individual 
stars, and therefore, not accounting for this effect will result in older ages
and lower star formation rates calculated from MS and BHeB stars.

\section{Tip of the Red Giant Branch Distance}
\label{S4}

The tip of the first ascent red giant branch (TRGB) has been empirically shown
to be a useful distance indicator for resolved, metal-poor stellar populations
({Lee, Freedman, \& Madore 1993}\nocite{lee93}; {Madore \& Freedman 
1995}\nocite{madore95}; {Sakai, Madore, \& Freedman 1996}\nocite{sakai96}; 
{Bellazzini, Ferraro, \& Pancino 2001}\nocite{bellazzini01a}).  The relative 
insensitivity of this CMD feature to metallicity (see below) suggests that 
TRGB distances are robust and may be reliably applied when other primary 
indicators are not available. One assumption in applying the TRGB method is 
that the red stellar population is old (\gsim\ 2 Gyr); however, this assumption
is met in every galaxy near enough for its older stellar population to be 
resolved (see the more detailed discussion in \S~\ref{S7.1}).

In nearby galaxies, the agreement between the 
TRGB distance and the distance derived from variable stars is excellent 
({Sakai \etal\ 1996}\nocite{sakai96} compare the Cepheid distance with the
TRGB distance in Sextans\,A; {Dolphin \etal\ 2001}\nocite{dolphin01} compare
Cepheid and RR Lyrae distances with the TRGB estimate in IC\,1613).  In a 
more extensive comparison of the TRGB with various distance indicators, 
\citet{ferrarese00a} find that in all 13 galaxies included in their sample 
with both TRGB and Cepheid distance estimates, these two distance indicators
agree within the total respective errors.  These and other results place the 
TRGB distance indicator on a reliable observational footing as a 
well-calibrated Population\,II secondary distance indicator.

We calculate the location of the TRGB using only the stars at large radial 
distance (Region\,1 in Figure~\ref{figcap6}) from the center of NGC\,625.   
This increases our photometric depth and reduces pollution by crowded faint 
stars that may blend together to mimic red giants in the central regions of 
the galaxy.  This region is free from recent star star formation and contains 
essentially only stars in the RGB region of the CMD.  Figure~\ref{figcap8} 
shows the resulting I-band luminosity function of the recovered stars in this 
region, binned in 0.05 magnitude intervals.   Note the steep breaks in the 
I-band luminosity function at I $\approx$ 23 and I $\approx$ 24 seen in 
both Figure~\ref{figcap8} and the CMD of Region\,1 shown in 
Figure~\ref{figcap7}.  The break at I $\approx$ 23 is a textbook quality 
example of the {\it false} TRGB due to population\,I AGB stars 
\citep[c.f.,][]{saha95}, and provides a reminder why it is important to 
observe to more than 1 magnitude below the TRGB for a secure distance 
determination. 

In Figure~\ref{figcap8}, the I-band luminosity function discontinuity
at I $\approx$ 24 indicates the TRGB for NGC\,625.  
We adopt the value of m$_{TRGB}$ = 23.95$\pm$0.07, and assume zero 
internal extinction for these stars.  Applying 
the theoretical and empirical calibrations of the TRGB absolute magnitude 
\citep{lee93,madore95,bellazzini01a}, we use the relation M$_I$ = 
$-$4.0$\pm$0.10, which should be valid for the metallicity range $-$2.8 $\leq$
[Fe/H] $\leq$ $-$0.6.  This will most certainly encompass the metallicity of 
the old stellar population in NGC\,625.  From nebular spectroscopy of the 
major \HII\ regions, {SCM03b}\nocite{skillman03b} find a weighted mean 
metallicity of 12+log(O/H) = 8.14$\pm$0.02 ($\sim$ 28\%\ of Z$_{\odot}$, or 
Z = 0.006, adopting the solar oxygen abundance of {Allende Prieto, Lambert, 
\& Asplund 2001}\nocite{allendeprieto01}). This provides an upper limit on the 
metallicity of the old stellar population, so the above range is 
well-justified.  Applying the metallicity-color relation derived in 
\citet{dacosta90} and also used in \citet{bellazzini01a}, we find that the mean
color of the TRGB in NGC\,625 is \vmi$_0 =$ 1.39.  This then corresponds to a 
mean [Fe/H] at the TRGB of $-$2.02, and to an I-band absolute magnitude safely
within the above quoted range.  Thus, it is not necessary to apply a 
significant correction to the assumed TRGB value of M$_I$ = $-$4.0$\pm$0.10.
The distance of NGC\,625 is then found to be 3.89$\pm$0.22 Mpc, placing it on 
the distant side of the three-dimensional structure of the Sculptor Group.  This 
is significantly more distant than the estimate of 2.7 Mpc by 
\citet{karachentsev03b} using a Tully-Fisher relation, although they note 
that this is an uncertain estimate from their expanded radial velocity $-$ 
distance relation.

\section{Stellar Populations, Evolutionary \& Synthetic Models}
\label{S5}
\subsection{Identification of Stellar Populations and Comparison to Stellar 
Evolution Models}
\label{S5.1}

The derivation of quantitative star formation histories depends
upon comparison of observed data with theoretical isochrones describing the
evolution of stars of different masses.  Here, we use the stellar evolution
models of the Padua group ({Bertelli \etal\ 1994}\nocite{bertelli94};
hereafter B94).  In order to conduct a first assessment, a representative 
value for the stellar metallicity of the dominant stellar population is
required.  For this, we rely upon two different lines of evidence to estimate the
stellar metallicity range in NGC\,625.  First, as previously mentioned, the 
mean nebular abundance in NGC\,625 is found to be 12+log(O/H) = 8.14$\pm$0.02 
{(SCM03b)}\nocite{skillman03b}.  Of course, the older stellar population should
have a metallicity lower than the current ISM.  Here, we adopt the Z=0.004 
isochrones of {B94}\nocite{bertelli94}.  We note that these isochrones
fit data on the stellar content of NGC\,6822 (which has similar 
global characteristics and nebular abundance to NGC\,625) quite well 
\citep[see, e.g.,][]{gallart94}.

This choice of metallicity is supported by the presence of the very red, 
extended AGB (see
\S~\ref{S7.2}) which generally only appears in stellar evolution models with Z 
$\gsim$ 0.001.  Furthermore, the MS and BHeB stars align better with the Z = 
0.004 models than with the Z = 0.001 models.  For these reasons, we adopt the Z
= 0.004 isochrones for the remainder of this work, and use them to determine, 
quantitatively in the case of the young stars and qualitatively in the case of 
the older stars, information about the recent star formation history of 
NGC\,625.  These isochrones are overlaid on the recovered photometry in 
Figure~\ref{figcap9}.  Note the good agreement of the 1-6 Gyr isochrones with 
the red, extended AGB stars detected at \vmi\ colors $>$ 1.5.

The photometry was next separated into four regions (MS, BHeB, RGB, \& AGB) by
placing selection polygons into the observed CMD.  These regions are based on 
the Z = 0.004 theoretical stellar isochrones discussed previously.  For the
young MS stars, the width of this regions was found by calculating 
the average error of the color index as a function of magnitude for stars 
with \vmi\ color index $<$ 1.0.  For the BHeB stars, a detailed simulation was 
performed, including the effects of differential extinction, to determine the 
locus of points which make up the BHeB sequence.  This model is described in 
detail in \S~\ref{S5.2}.  Note that the constant-color red edge of the MS
selection region is driven by the need to assure exclusive populations between
MS and BHeB stars; since the MS stars are not used for quantitative calculations
of the star formation history, this will only affect the derived spatial 
distribution of MS stars.  Selection regions for the 
older stars were created by hand. In an attempt to only retain the RGB stars, 
our RGB selection region was guided by the photometry for Regions 1 \&\ 2 (see
Figure~\ref{figcap7}), which contain predominantly old stars that are 
well-separated from the main disk of the galaxy.  Finally, the selection 
region for the AGB stars was guided by noting the position of the AGB in the 
aforementioned models, and by selecting stars which follow the locus of points
for older stars (ages $\sim$ 1-6 Gyr) delineated therein.  This region then 
effectively avoids the densely populated red region of the RGB, which, being 
near the edge of our photometry, will have relatively high photometric errors. 

\subsection{Modeling Differential Extinction in Young Stellar Populations}
\label{S5.2}

The differential extinction, which blurs the MS and BHeB stars in the CMD, 
makes our photometric separation of these populations difficult.  To better 
gauge the severity of this effect, in Figure~\ref{figcap10} we show a closer 
view of the Region\,6 photometry.  Overlaid are the zero-age MS and
the sequence showing the blue extent of the core helium burning stars.  Note 
that the wide distribution in color of the young stars argues for pronounced 
extinction along some (but not necessarily all) sightlines.  Coupled with our 
narrowband imaging, where small-scale variations in the extinction were 
measured over distances as small as tens of parsecs, we conclude that
differential extinction is non-negligible.  This effect must be accurately 
accounted for before a realistic separation of these sequences (which are 
separated by $\sim$ 0.3 magnitudes in \vmi) can be made.

To explore the effects of differential extinction in detail, we performed 
simulations of star formation events using the aforementioned
stellar evolution models of {B94}\nocite{bertelli94}.  Here, an initial star
formation rate and event duration are specified.  Stars are then ``created''
according to a power-law distribution in mass that is governed by the IMF (here,
a Salpeter IMF was applied), and then assigned an age (depending on the length 
of the star formation event).  The temporal resolution is set at 10$^5$ years, 
assuring that the short-lived stages of stellar evolution considered here 
(RGB, red helium burning (RHeB) and BHeB stages) are extractable.  Note that 
MS luminosity evolution is explicitly accounted for.

These stars, now having an age and a mass, are converted into the V vs. \vmi\
plane by interpolation over the model grid.  Comparing the ages of these stars 
to model ages of various stellar evolution phases, one can 
then uniquely determine its position in the CMD.  Furthermore, the temporal 
resolution assures that one can extract synthetic stars in the stages of interest
here, namely the MS and BHeB stars.  The results of this interpolation produce
synthetic photometry representing various stellar populations, including 
evolving MS stars, red giants, BHeB and RHeB stars.

After this photometry is created, a simple random number generating routine is 
used to apply first differential extinction and then photometric errors to the 
synthetic data.   Based on our extinction measurements, a flat distribution in 
differential extinction over the range of 0.0 $<$ A$_V$ $<$ 0.6 is assumed.
Using least-squares 
fitting to these points in the V vs. \vmi\ plane, the location of the BHeB
selection region can then be derived.  As will be described in \S~\ref{S6.3},
these simulations are also used to self-consistently measure the degree of 
contamination of the BHeB selection regions by reddened MS stars.  

The result of this photometric selection process is to produce different
visualizations of the distribution of stars of various ages and evolutionary 
phases throughout the galaxy. In Figure~\ref{figcap11} we show the selection 
regions used to extract the populations (see above for further details).
Note for the BHeB selection region that the diagonal appearance of the region 
boundaries comes from stars moving down and to the right in the V vs. \vmi\ 
plane as a result of differential extinction (see also \S~\ref{S6.3}).
Figure~\ref{figcap12} presents density plots of the various stellar
populations, showing the number of detected \stpkpc.  We detect a very clear age
segregation (i.e., a stellar population gradient) in NGC\,625: the young stars 
(MS, BHeB) are highly 
concentrated toward the disk, and align well with the locations of star 
formation (as evidenced by \halpha\ emission; see \S~\ref{S6.1}) and 
high-column density \HI\ emission.  Note, however, that the largest star 
formation region (NGC\,625\,A) does not coincide exactly with the position 
of the highest column density \HI\ emission {(CCF00)}\nocite{cote00}, but is 
offset to the east by roughly 300 pc.  In contrast to the young stars, as the 
age of the stellar population increases (AGB stars, RGB stars), the 
distribution of stars becomes much more diffuse and uniform throughout the 
galaxy.  Each of these facets of the stellar populations of NGC\,625 will be 
discussed in the sections that follow.

\section{The Young Stars and Recent Star Formation History}
\label{S6}

Our WFPC2 data allow us to study both the newly formed stars and the older
stellar population of NGC\,625.   While the orientation of the WFPC2 field of 
view does not encompass all of the galaxy (see Figures~\ref{figcap1} and 
\ref{figcap2}), we do have a representative view of the central and eastern 
half of the main disk (including all of the major star formation regions), as 
well as the surrounding halo region.  In this section we discuss the recent 
star formation history of NGC\,625 that can be discerned from our data.  We 
discuss both the MS stars and BHeB stars in this context, and use the latter 
to derive a simple model of the recent star formation history of NGC\,625.

\subsection{The Main Sequence Stars}
\label{S6.1}

Figure~\ref{figcap10} demonstrates that MS stars with ages approaching 20 
Myr can be identified in our photometry (with the previously noted ambiguity 
due to differential reddening).  The spatial distribution of young 
(i.e., blue plume) stars is highly concentrated toward the central disk of 
NGC\,625. As Figures~\ref{figcap12}(a) and \ref{figcap12}(b) reveal, the young 
stars are predominantly located near the large \HII\ complexes, with the bulk 
of both populations highly concentrated near the largest star formation region 
(NGC\,625\,A, see Figure~\ref{figcap3}).  Not all of the MS stars are 
associated with detected \HII\ regions, however; many are located quite far 
away from the lowest levels of detected \halpha\ emission.  To better 
visualize the spatial coincidence of the detected MS stars and the locations 
of the major active star formation regions (i.e., those demonstrating \halpha\
emission), we plot in Figure~\ref{figcap13} the locations of all of the stars 
identified as young MS stars (M$_V$ $<$ $-$2.5; see Figure\,11) which lie in 
the central region compared with contours of the \halpha\ emission.  The most 
luminous regions, NGC\,625\,A and B, are both associated with many detected MS
stars, while NGC\,625\,C is associated with fewer recovered MS stars, in line 
with expectations due to its lower \halpha\ flux.

The spatial correlation between MS stars and \halpha\ emission is biased by
the low \halpha\ surface brightness sensitivity of the HST (relative to lower
angular resolution ground based imaging) resulting in non-detections of known
\HII\ regions.  While the undetected emission is a small fraction of the total
\halpha\ emission in NGC 625, it does affect the comparison of MS star 
locations with \halpha\ emission. Figure~\ref{figcap1} of 
{SCM03a}\nocite{skillman03a} reveals a large number of smaller \HII\ regions 
throughout the disk of NGC\,625 which we do not detect here, and which most 
likely are associated with the MS stars which we detect in the present 
photometric analysis.  Low-surface brightness \halpha\ emission is present 
throughout (and slightly beyond) the field of view shown in 
Figure~\ref{figcap13}.  The fact that some MS stars are associated with 
lower-surface brightness \HII\ regions (or none at all) suggests that recent 
star formation has been widespread throughout the central area, and not 
necessarily confined to the large associations detected here. 

\subsection{The Blue Helium Burning Stars}
\label{S6.2}

Figure~\ref{figcap10} demonstrates that BHeB stars with ages approaching 
200 Myr can be identified in our photometry (with the previously noted 
ambiguities due to differential reddening and the confusion with 
reddened MS stars).  Like the MS stars, the distribution of BHeB stars is 
concentrated toward the disk, although slightly more extended.  This could 
either represent more extended prior star formation or diffusion from
natal formation sites, although the former seems more likely, based on typical 
stellar diffusion velocities.  Interestingly, an extension of young stars to 
the SE is detected prominently in the BHeB stars (see Figure~\ref{figcap12}b).  
This region shows that recent star formation was not confined to just the central,
high-\HI\ column density regions of NGC\,625.  The recent enhanced star formation 
appears to have been more of a distributed as opposed to concentrated phenomenon.

\subsection{A Preliminary Model for the Recent Star Formation History of 
NGC\,625}
\label{S6.3}

As discussed in \S~\ref{S5}, the effects of differential extinction are larger 
than the theoretical separation of the MS and BHeB stars in the blue plume.  
Thus, a detailed recent star formation history derived directly from the BHeB 
luminosity function (e.g., {Dohm-Palmer \etal\ 1997b}\nocite{dohmpalmer97b}, 
{2002})\nocite{dohmpalmer02a} is not straightforward. On the other hand, given 
the uncertainties introduced by the differential reddening and the 
increased photometric depth that would be required to reach even moderate ages
(at least 1 full magnitude deeper to reach 500 Myr; see Figure~\ref{figcap10}),
we do not feel that a Monte Carlo simulation of the blue plume presents a 
promising technique for deriving a recent star formation history either.  
Nonetheless, we are confident that the present observations carry information 
relevant to important questions concerning recent star formation in NGC\,625.
Specifically, is the recent star formation best described as an instantaneous 
burst, as inferred, for example, in BCDs with strong Wolf-Rayet (W-R) features 
\citep{kunth81}?  Recent studies, using updated stellar models, find ages for 
starbursts in so-called ``W-R galaxies'' of 3$-$6 Myr with limits on burst 
durations of 2$-$4 Myr {(Schaerer, Contini, \& Kunth 
1999a)}\nocite{schaerer99a}.  Alternatively, is the enhanced star formation 
better described as long lived with a lifetime in excess of 10 Myr? 
Additionally, has the recent star formation been mainly confined to the very 
center of NGC\,625 (or a single star forming region), or has it been 
distributed throughout the central disk?

In order to answer these relatively simple questions, we will examine the 
most luminous BHeB stars identified in Figure~\ref{figcap11}.  While there is 
some ``pollution'' by reddened MS stars (see below) and a fair degree of 
uncertainty in magnitude due to the differential reddening, if we examine the 
BHeB luminosity function in a {\it statistical} sense and at low temporal 
resolution (i.e., bin sizes larger than the size of the measured typical 
reddening vector, 0.5 magnitudes in the V-band), we should be able to construct
first-order answers to the above questions.  Indeed, we note that, even in 
nearby low-metallicity systems where differential extinction is expected to be 
minimal, the observational separation of the MS and BHeB is not perfect, and a 
statistical separation must be employed (e.g., GR\,8, {Dohm-Palmer \etal\ 
1998}\nocite{dohmpalmer98}; Sextans\,A, {Dohm-Palmer \etal\ 
2002}\nocite{dohmpalmer02a}).  Since these investigations have produced 
reliable star formation histories (note the agreement of the recent SFH 
derived from BHeB and MS stars in Sextans\,A, for example), we conclude that 
the separation of the MS and BHeB sequences in a statistical sense is a robust 
technique, even considering the strength of differential extinction in this 
system.

As shown in Figure~\ref{figcap10}, incompleteness corrections for the BHeB 
stars should be small for M$_V$ $<$ $-$4, corresponding to unreddened ages 
$\lsim$ 100 Myr.  These relatively massive (M $\gsim$ 5.2 \msun), luminous 
BHeB stars provide temporal star formation resolution of $\sim$ 25 Myr in 
this age interval.  Since the luminosity of a BHeB star corresponds to a 
single age \citep[see, e.g., discussions in][]{dohmpalmer97b,dohmpalmer02a}, 
these blue supergiant stars are excellent probes of the recent star formation 
history.  Thus, by counting the numbers of such stars, and applying conversion 
factors that account for the IMF and amount of time spent in the BHeB phase, 
one can glean temporal and spatial information about the recent star formation 
activity in a galaxy.  

We separate the BHeB photometry into four 
magnitude bins, which correspond to age regions based on the theoretical 
isochrones of {B94}\nocite{bertelli94}: 0-25 Myr, 25-50 Myr, 50-75 Myr, and 
75-100 Myr.  Figure~\ref{figcap11} shows these population regions separated 
in the BHeB selection box by dotted lines.  Note that each selection region
has boundaries that follow the slope of the reddening vector (A$_V$ $=$ 0.6;
see \S~\ref{S5.2}).  Note also that each region is above the 80\%\ completeness
line as derived from artificial star tests, and hence no corrections are 
necessary.  

In Figure~\ref{figcap14} we show the spatial distribution of the 
BHeB stars in these four magnitude bins.  The clear separation into 
identifiable star forming regions in the youngest three epochs indicates that 
the coarse binning has overcome some of the ``noise'' introduced by the 
differential reddening.  In particular, the brightest stellar association today
(associated with the giant \HII\ region) is seen to be relatively young, while 
the stellar association to the east appears to be roughly 20-40 Myr older.  
Furthermore, comparing the highest contour for the 0-25 Myr population (shown
in white in Figure~\ref{figcap14}) with previous star formation peaks, it is
clear that star formation has been moving throughout the disk during this 
time period.  Taking these data at face value, it appears that a prior, stronger 
burst pervaded the disk 50-100 Myr ago.  

Formally, analysis of the coarse BHeB luminosity function implies a large and 
declining star formation rate over the last 100 Myr.  The relative star
formation rates (originally in units of \msun\,yr$^{-1}$, then 
normalized to unit intensity) over the last 100 Myr are 1.0$\pm$0.38 (0-25 Myr),
1.3$\pm$0.38 (25-50 Myr), and 5.0$\pm$1.3 (50-75 Myr) 5.0$\pm$3.8 (75-100 Myr). 
Errors here reflect Poisson statistics on the number of stars detected in each 
age bin, as well as average contamination factors as derived from the 
simulations described below.  While we defer more detailed conclusions about the 
past star formation history of NGC\,625 to future investigations where multicolor 
photometry \citep[e.g., U,\,B,\,V,\,I; see][]{romaniello02} can be used to 
ascertain detailed spatial information about the dust and to derive line-of-sight 
corrections for it, this simple model of the recent evolution of NGC\,625 should 
not be dramatically altered.  We note that these star formation rates are lower
limits, as there will undoubtedly be BHeB stars that suffer little or no 
differential extinction but undergo photometric errors carrying them blueward of
the selection regions here; however, this effect should be independent of 
magnitude and hence will not drastically alter the {\it relative} star formation
rates for each coarse age bin. 
We conclude that the average SFR from 100-50 Myr ago was a factor of $\sim$ 5 
higher than the average SFR from 50 Myr to the present epoch.  This suggests 
that we may be witnessing the final stages of the extended star formation 
episode in NGC\,625, and that this system may be rightly qualified as a 
``post-starburst'' galaxy.

The contamination of the BHeB population by reddened MS stars is measured 
self-consistently using the simulations described in \S~\ref{S5.2}.  Here, 
the recovered star formation rates quoted above are input into 100 realizations
of our synthetic model, which includes differential extinction (0.0 $<$ A$_V$ $<$ 
0.6 magnitudes).  Each model iteration provides an independent output
photometry set, and hence an independent measure of the contamination of the 
four segments of the BHeB selection region by reddened MS stars.  The average 
contamination factors from all 100 iterations (for each age bin) are then added 
into the total error budget of the star formation rates derived from the BHeB 
stars as described above.  While some level of ambiguity may remain due to 
the choice of isochrones, we again emphasize that the {\it relative} star 
formation rates, with errors, should present as accurate a picture of the recent 
star formation history of this system as is possible from these data.  This 
simple analysis implies that the burst of star formation in NGC\,625 has a 
relatively long duration (of order 100 Myr) and has been widespread throughout 
the disk.

Such a scenario is consistent with the other observations of this system which 
were briefly discussed in \S~\ref{S1}.  In particular, \citet{bomans98} find 
diffuse soft (ROSAT 0.1 - 2.4 keV) x-ray emission above the northern side of
the disk.  The presence of this hot ($\sim$ 10$^6$ K) gas at large distances
from the current or recent star formation complexes ($\sim$ 1 kpc North and 
East of complex NGC\,625\,A) suggests that an active outflow has been at work 
in this system in recent times.  For example, if an average outflow velocity of 
100 \kms\ is assumed, only $\sim$ 10 Myr would be needed; even if the average 
velocity is an order of magnitude smaller, the extended star formation event 
forwarded above could easily have expelled the gas to this distance.  However, 
in the spectroscopic investigation of the 
ionized gas by {MMHS}\nocite{marlowe97}, no direct evidence of an active outflow
from the current major star formation complex is found. Given that the 
star formation rate has declined during the last 50 Myr, it appears that this 
is an ideal system in which to investigate the physical processes that 
terminate or slow hot gas outflows from major star formation complexes. Further
x-ray imaging is under analysis \citep{bomans03}, and when combined with deeper 
multi-color photometry of the starburst regions, we should be able to address 
this interesting and important issue in greater detail.

The most important conclusion of this star formation history analysis is that
NGC\,625 appears to have sustained a heightened star formation rate for an 
exteneded period of time (i.e., \gsim\ 10 Myr).  With the aide of this modeling 
approach, we can easily discern between bursts of star formation that are of 
short ($\sim$ 5 Myr) or long ($\sim$ 50 Myr) duration.  By comparing our data 
to various model realizations, we reject any model which does not sustain 
massive star formation long enough to populate the BHeB region of the CMD.  To 
visualize this, we present in Figure~\ref{figcap15} a comparison of two 
synthetic bursts of star formation; one of short duration (5 Myr), and one of 
long duration (50 Myr).  It is clear that the short-duration burst fails to 
populate the BHeB region of the CMD, while the longer-duration burst is more 
successful at placing appreciable numbers of stars in this region.  Note also 
the presence of numerous stars in the color region \vmi\ $>$ 0.1 in the empirical
photometry, and the pronounced dearth of stars in this location in the 
short-duration burst model.  From these tests we conclude that the star formation
in NGC\,625 is of extended duration compared with most models of starbursts
(durations $\sim$ 3$-$6 Myr), and that the presence of spectroscopic W-R features 
does not rule out a longer-duration star formation event.  

\section{The Older Stellar Population}
\label{S7}

Here, we analyze the older stellar populations in NGC\,625, namely the 
intermediate and old age AGB and RGB stars.  While the depth of photometry 
here does not allow the derivation of a quantitative star formation history 
for intermediate and old ages, we can take a qualitative look at the past 
evolution of this system.  

\subsection{The Red Giant Branch}
\label{S7.1}

To date, all nearby low-metallicity
dwarfs have been shown to contain some fraction of an older stellar population,
although in some cases it may be very small \citep[e.g., the very 
low-metallicity Local Group dwarf Leo\,A,][]{tolstoy98,dolphin02c}.  This red 
population
appears to be present in all galaxies that are near enough to allow it to be 
resolved.  The spatial distribution of RGB stars is expected to trace the general 
morphology of a galaxy at the time when the dominant component of its old 
stellar population formed. By studying the spatial distribution of the older 
stars, we may hope to attain a basic understanding of the past star formation 
history of a galaxy.  Of course, stellar evolution models become moderately 
degenerate in age and color on the RGB after $\sim$ 1 Gyr (i.e., temporal 
resolution becomes much more difficult than for younger stars in the blue 
plume), so using such features for quantitative analysis of the past star 
formation history requires both deep photometry and sophisticated modeling 
\citep[][and references therein]{dolphin02b}.  Due to the distance of NGC\,625,
such an analysis of the RGB is not feasible.  Rather, we use these stars to 
demonstrate the presence of an old stellar population, and use their spatial 
distribution to understand various facets of the galaxy's history in a 
qualitative sense.

A comparison of the RGB stars with the overall \HI\ distribution of NGC\,625 
demonstrates good agreement.  In the analysis of VLA observations of this 
southern target, {CCF00}\nocite{cote00} were unable to unambiguously model the
\HI\ mass distribution or to derive a rotation curve. This is due to the highly
disturbed and complex \HI\ distribution found in this galaxy, with 
multi-peaked \HI\ emission and apparent large-scale rotation about the major
axis rather than the minor axis.  In a large-scale sense, however, the 
high-column density \HI\ distribution follows the stellar disk of this galaxy 
quite well, with a nearly east-west major axis that appears to align with the 
distribution of RGB stars (at least throughout the stellar disk of the 
galaxy).  This agrees with the position angle of the major axis (92\degree) 
derived by {MMHS}\nocite{marlowe97}.  Recall from \S~\ref{S3.2} that we applied 
elliptical selection regions of axial ratio 2.8:1 to isolate crowding effects 
in our photometry.  These regions closely trace the isophotal contours of the
stars, in particular those of the RGB stars in the outer regions of the disk 
and in the halo.  The photometric center of these regions (and of the galaxy) 
falls very near the center of the stellar disk, nearly equidistant between the
\HII\ complexes NGC\,625\,B and NGC\,625\,C  ($\alpha,\delta$ (J2000) = 
01:35:05.33, -41:26:09.9).  The overall agreement of the \HI\ and RGB 
distributions suggests that stars have been forming in the present disk-like 
morphology for a very long time.

\subsection{The Asymptotic Giant Branch}
\label{S7.2}

The AGB selection region shown in Figure~\ref{figcap11} isolates stars that
are extremely red in color (\vmi\ $>$ 1.5 and extending to 3).  As this 
feature is not predominant in all galaxies (see below), its unexpected 
strength warranted a careful exploration of the errors associated with the 
photometry of these particular stars.  The extremely red colors of some of 
these stars must, of course, be the result of errors in the \vmi\ color index.
Indeed, as the limiting magnitude uncertainty is decreased, this extension 
becomes smaller and less-populated.  However, even at the very low error limit
of $\sigma$ $<$ 0.1 magnitudes in both I and V, the feature is still detected 
prominently and is therefore robust.  Note also that, in Figure~\ref{figcap7},
most of these AGB stars lie above the 80\%\ completeness lines, suggesting 
only minor corrections for such effects.

The spatial distribution of these red AGB stars varies relatively smoothly, 
much like the distribution found for the RGB stars (see \S~\ref{S7.1}).
However, it is confined closer to the disk of the galaxy than are the red 
giant stars, again reinforcing the segregation of the various stellar 
populations based on age.  The apparent under-density of AGB stars near the 
largest star formation region is most likely not real, but is caused by the 
lower spatial resolution of the WF chip having difficulty with the highly
crowded stars in this region, and the relatively under-luminous AGB and RGB 
stars being overcome by the more massive and brighter MS and BHeB 
stars.

The ratio of RGB stars to AGB stars changes as a function of galactic radius. 
In Table~\ref{t6} we show this ratio as a function of region (see 
Figure~\ref{figcap6}).  The ratio falls from values near 40 in the outer 
regions, which are dominated by RGB stars, to values less than 10 in the 
central regions, where AGB stars are more prominent.  Note that, in the 
central sections of the galaxy, our sensitivity to these comparatively 
under-luminous stars drops, as they can easily be overwhelmed by brighter, 
younger stars.  As previously noted, the shape of the RGB and
AGB selection regions was chosen to follow the empirical distribution of stars
in these regions, as well as the general position and shape of the theoretical
isochrones.  While the numerical results will be slightly dependent upon the
choice of the shapes of these regions, the overall result is quite robust, in
that we detect a decreasing number ratio of RGB to AGB stars as a function of
decreasing galactocentric distance.  We interpret this as verification of the
stellar population gradient in NGC\,625.   The presence of a fairly 
extended old stellar halo in NGC\,625 argues for active star formation in the 
distant past \citep[c.f.,][]{minniti97}.  Since AGB and RGB stars both have 
large and overlapping ranges in age, the AGB to RGB ratio cannot be converted 
directly into an unambiguous star formation history constraint.  However, 
under the simple assumption that the AGB population is, on average, younger 
than the RGB population, the gradual increase in the AGB to RGB ratio with 
decreasing radius suggests a slow decrease in the active area of star 
formation in this dwarf galaxy.  Of course, this ignores possible metallicity 
dependences in the AGB luminosity function and other ambiguities, but the ratio
may be indicating something both simple and fundamental about the history of
star formation in NGC\,625.

\placetable{t6}

According to the stellar evolution models of {B94}\nocite{bertelli94}, at the 
metallicity of NGC\,625, this red AGB region of the CMD is populated only by 
AGB stars with ages of a few Gyrs that are relatively metal-rich (Z $\gsim$
0.001 Z$_{\odot}$).  This suggests that these stars formed from gas which 
underwent previous metal enrichment by an earlier generation of stars.  The 
strong RGB population is most likely composed of the low-mass siblings of the 
stars that enriched this gas, implying that star formation was ongoing early 
in the history of NGC\,625, approaching a Hubble time ago.  The detection of 
this morphology agrees with theoretical isochrones for metallicities higher 
than 0.001 Z$_{\odot}$ {(B94)}\nocite{bertelli94}.  Note the general agreement 
of the extended metal-rich isochrones with the data in Figure~\ref{figcap9}. 
These models demonstrate that the bulk of the AGB stars have ages in the range of 
2-6 Gyr.  The definitive detection of stars which are predominantly old (RGB 
stars, ages $\sim$ 10 Gyr), middle age (these AGB stars), and young (see 
\S~\ref{S6}) indicates that NGC\,625 has been forming stars throughout the 
entire lifetime of the galaxy (although, for ages $\gsim$ 1 Gyr, the rate of 
star formation remains unconstrained). 

A similar AGB is seen in several galaxies where deep HST photometry has been 
extracted.  While this feature is weak in most cases, perusing the \vmi\ 
colors of extracted photometry for some local star-forming galaxies, one finds
what may be a similar population of stars in Sextans\,A 
\citep{dohmpalmer97b,dohmpalmer02a}, VII\,Zw\,403 
\citep{lynds98,schulteladbeck99a}, UGCA\,290 \citep{crone02}, and NGC\,1705 
\citep{tosi01}.  However, undoubtedly the strongest detection of this CMD 
feature is found in the Local Group dIrr galaxy NGC\,6822.  Therein, 
\citet{gallart94} and \cite*{gallart96a} find a densely populated, red AGB 
that occupies the color range 0.9 $\leq$ (V$-$R)$_0$ $\leq$ 1.7.  The authors 
interpret this ``red tail'' as evidence for an intermediate to old stellar 
population.  The red tail found here seems to be quite similar in color and 
absolute magnitude to that found in NGC\,6822.  While the above sample surely 
is not complete (nor is it exclusive), it serves to demonstrate that, in many 
cases, such a population of intermediate-age stars is a quite common component
of a galaxy's stellar content, and hence, also of its star formation history.  

\section{The Burst Triggering Mechanism}
\label{S8}

The mechanisms and physical processes by which star formation begins and is 
sustained in galaxies are important, but by no means fully understood.  In 
dwarf galaxies in particular, the questions of how star formation begins and 
proceeds are especially difficult to address, as these objects have no known 
internal large-scale mechanisms to regularly prompt the process of star 
formation (e.g., spiral density waves).  Evidence is mounting for the importance of 
environmental factors which may play a crucial role in the evolution of some 
of these actively star-forming galaxies, requiring an understanding of not 
only the detailed nature of the ISM but also the nature of the local 
conditions in which a galaxy is found (see, e.g., 
{SCM03a}\nocite{skillman03a}).  We discuss here some possible scenarios for 
the evolution of NGC\,625, an apparently isolated galaxy.  Our spatially 
resolved stellar population information allows us to examine important 
characteritics  such as the correlation between neutral gas and newly formed 
stars, the size and distribution of star formation regions, and the timescales 
over which such regions were active in the past.

NGC\,625 sits near the edge of the three-dimensional structure of the Sculptor
Group (extended along the line of sight; see \S~\ref{S4}).  It is natural to
posit that a nearby Sculptor member may have triggered the rigorous star
formation episode that the galaxy is undergoing.  The nearest luminous galaxy 
is ESO\,245-005, with a projected (minimum) separation of 190 kpc. ESO\,245-005
differs in Galactocentric radial velocity by only 19 \kms.  This corresponds to
a deprojected distance of 580 kpc (using the TRGB distance for NGC\,625, and 
the TRGB distance estimate of 4.43$\pm$0.45 Mpc for ESO\,245-005 from
{Karachentsev \etal\ 2003}\nocite{karachentsev03b}). Even assuming the 
minimum separation of 190 kpc, a relative velocity in excess of 1000 \kms\ 
would be required for a near passage ($\le$ 50 kpc, comparable to the distance
of the Magellanic Clouds) within the recent past ($\le$ 100 Myr).  This 
suggests that either dwarf galaxies can trigger starbursts via interactions
from distances of order 200 kpc or that ESO\,245-005 is not responsible for the
present burst of star formation.  Note that ESO\,245-005 does exhibit a 
disturbed \HI\ velocity field {(CCF00)}\nocite{cote00},  but one possible 
explanation is that ESO\,245-005 is undergoing a merging of two dwarf systems. 
Alternatively, it has been suggested that the \HII\ regions concentrated at the
ends of the bar are indicative of an interaction \citep[c.f.,][]{miller96}. 

{CCF00}\nocite{cote00} posit that NGC\,625 may have undergone a merger 
event recently, but with a non-luminous (or now-accreted) companion (perhaps an
\HI\ cloud).  Indeed, there exist many \HI\ clouds coincident with the Sculptor
Group, but they display a wide velocity dispersion about the systemic velocity
of the group, and none are coincident in both position and velocity with 
NGC\,625 \citep[see][]{putman02}. If a merger between NGC\,625 and an
\HI\ cloud occurred prior to the current star formation event, the elevated gas
densities may have been sufficient to trigger the current burst.  In addition, 
the observed abundance ratios ({SCM03b}\nocite{skillman03b}; see further 
discussion below) would be consistent with chemodynamical models of \HI\ infall
events \citep{hensler99}. Similarly, a merger event cannot be ruled out 
for ESO\,245-005, although an interaction between the two galaxies seems 
unlikely as the catalyst of the current star formation events.  Of course, the 
galaxies may be interacting on longer timescales, and tidal effects cannot be 
ruled out completely as having had an effect on the current evolutionary status
of the two galaxies (see {Taylor 1997}\nocite{taylor97} for a more complete 
study of the frequency of companions in such systems).

The highly disturbed \HI\ velocity field {(CCF00)}\nocite{cote00} and the 
confused \halpha\ kinematics {(MMHS)}\nocite{marlowe97} suggest a merger 
scenario.  We do see clues in the combination of our stellar data and the \HI\ 
which suggest that the main \HI\ and stellar disk of NGC\,625 has been warped. 
Recall the extension of young stars toward the southeast of the stellar disk; 
this area corresponds to relatively low-column density \HI\ gas, and suggests 
that the stars therein may have been perturbed gravitationally and carried away
from their formation sites.  It is then most intriguing that there also exists 
a low-column density \HI\ extension toward the northwest of the stellar disk 
(see Figure\,1 of {CCF00}\nocite{cote00}). The symmetry of these potential 
warps (although note that they are on very different physical scales) might 
suggest a tidal interaction with an unseen counterpart.  While we do not 
have optical data for this region, such an investigation might prove fruitful 
for discerning the nature of the apparent interaction which has left its 
signatures on both the stellar component, as well as the ionized and neutral 
gas kinematics, of NGC\,625.

If the recent burst is not the result of an encounter with a relatively 
massive companion, what other scenarios could explain the observed 
characteristics and star formation history? From chemical evolution arguments,
we might be able to gain some insight into the nature of the current burst of 
star formation.  {SCM03b}\nocite{skillman03b} find a high N/O ratio for the 
three \HII\ regions studied in NGC\,625 (weighted mean log(N/O) $= 
-$1.30\,$\pm$\,0.02; see their Table\,2).  This is a high value for N/O at this
O/H (see compilation in {Kobulnicky \& Skillman 1996}\nocite{kobulnicky96}), 
and could be interpreted as evidence for a long quiescent period preceding the 
current burst of star formation.  During this time, the intermediate mass stars
(likely the predominant producers of N) would have time to lose much of their N
to the ISM, making it appear comparatively N-rich \citep{garnett90}.  
\cite*{skillman97} find high N/O in Pegasus (log(N/O) $\sim -$1.25), and posit 
a similar sequence of events leading to the current (albeit very mild) star 
formation episode.  The HST imaging study of \citet{gallagher98} confirms that 
the global star formation rate in Pegasus has been low for the past few 100 
Myrs, suggesting that the delayed N mechanism is consistent with observations 
of this particular galaxy.  If this delayed N production scenario is valid for 
NGC\,625 as well, then perhaps the disrupted ionized and neutral gas velocity 
fields are the result of the current star formation episode.  That is, perhaps 
we are witnessing the in situ disruption of the disk of NGC\,625 by the ongoing
starburst activity.

Of course, since we lose temporal resolution of star formation activity after 
$\sim$ 0.1 Gyr, we cannot discern any scenario which would take such a timescale 
to leave its observational signature on the galaxy.  However, any consistent
scenario must account for the recent star formation (i.e., behavior on 
timescales of Myr), as well as explain the disrupted velocity fields, the 
large-scale \HI\ distribution, the displaced young stars toward the southeast
region of the galaxy, and the overall heightened star formation rate throughout
the galaxy.  

\section{Comparison with Similar Starbursts}
\label{S9}

\subsection{NGC\,625 As A Blue Compact Dwarf Galaxy}
\label{S9.1}

The class of star-forming galaxies known as BCDs includes low-luminosity 
systems that have concentrated and rigorous star formation.  The first 
systematic investigation of these galaxies was given in \citet{thuan81}, where 
various criteria (luminosity, spectra and physical size) were used to 
differentiate between these systems and other low-mass galaxies.  Since then, 
numerous investigations have shown that while these galaxies share certain 
properties, the class remains somewhat heterogeneous with respect to 
morphology and other properties (e.g., {Loose \& Thuan 1986}\nocite{loose86}; 
{Papaderos \etal\ 1996a}\nocite{papaderos96a},{b}\nocite{papaderos96b};
{Doublier \etal\ 1997}\nocite{doublier97}; {Doublier, Caulet, \& Comte 
1999}\nocite{doublier99}; {Cair{\' o}s \etal\ 
2001a}\nocite{cairos01a},{b}\nocite{cairos01b}).  \citet{hopkins02} find a 
wide range in current star formation rates in such galaxies spanning 
nearly five orders of magnitude, with a median value of $\sim$ 0.3 
\msun\,yr$^{-1}$.

More recently, {Gil de Paz, Madore, \& Pevunova (2003)}\nocite{gildepaz03} have
produced a statistically large (although not complete) sample of nearby BCDs, 
with the aim of understanding the properties of this class of galaxy.  These 
authors postulate new selection criteria for classification as a BCD, that are
somewhat different and more quantitative than the previous standard of 
\citet{thuan81}. \citet{gildepaz03} suggest that criteria involving the K-band
luminosity (M$_K$ $>$ $-$21), peak surface brightness ($\mu_{B,peak} <$ 
22\,mag\,arcsec$^{-2}$), and color at the peak surface brightness 
($\mu_{B,peak} - \mu_{R,peak}$ {\lsim} 1) be satisfied in order to classify a 
galaxy as a BCD.  While, at present, few dwarf galaxies have measurements of their 
K-band luminosities, this work provides useful transformations based on the 
more common (B$-$R) color index.  Using this new definition of BCD, 
\citet{gildepaz03} note that some traditional BCD galaxies no longer fit these
new criteria (notably, II\,Zw\,33 and Tol\,1924-416), but for the most part, 
the galaxies selected with these new benchmarks agree with those selected by 
\citet{thuan81}. The advantage lies in the more quantifiable nature of the 
selection process.

Given these new selection criteria, does NGC\,625 qualify as a bona fide BCD 
galaxy?  Applying the broadband colors found by \citet[][assuming that the 
``core'' region therein is representative of the peak surface brightness, and 
that surface brightness colors scale as do broad-band colors]{marlowe97}, 
NGC\,625 satisfies each of the criteria of \citet{gildepaz03}.  The system 
also satisifies the criteria of \citet{thuan81}.  Finally, as discussed in 
{SCM03b}\nocite{skillman03b} and in \S~\ref{S8}, NGC\,625 demonstrates 
spectroscopic properties that are seen predominantly in blue compact dwarf 
galaxies.  Thus, NGC\,625 should be added to BCD surveys.
With this in mind, it is insightful to compare the properties of NGC\,625 
with those found in other nearby, well-resolved starburst and BCD galaxies.  
In Table~\ref{t7}, we present such a comparison, highlighting important 
characteristics of both the galaxy and its contained starburst region.  
Note that this list is neither exhaustive nor complete, but rather draws 
attention to the most commonly-studied systems within $\sim$ 5 Mpc.

\placetable{t7}

Table~\ref{t7} demonstrates a wide variety of properties among these 
selected starburst galaxies.  Note, for example, the large range in L$_X$/SFR
and L$_{100\,\mu{m}}$/SFR between otherwise similar galaxies.  Systems showing
comparable masses, metallicities and current star formation rates show a large
disparity in x-ray and FIR luminosities (compare NGC\,1569 and NGC\,1705).
Similarly, systems showing similar x-ray luminosities, masses, and other 
characteristics differ in global SFR (compare NGC\,625 and NGC\,1705).  Further
similarities can be drawn from resolved stellar population studies of these 
systems; for example, the old stellar populations in NGC\,625 and VII\,Zw\,403
are both uniformly distributed (see Figure~\ref{figcap12}d and 
{Schulte-Ladbeck, Crone, \& Hopp 1998}\nocite{schulteladbeck98}), and both show
clear evidence of stellar population gradients.   Yet, their current star 
formation rates, FIR and x-ray luminosities differ markedly and somewhat 
non-intuitively (note that NGC\,625 has a higher current SFR and FIR 
luminosity, but VII\,Zw\,403 outshines NGC\,625 in x-rays).  This collection of
data argues for the importance of detailed studies of these systems to 
decipher the nature of star formation in active low-mass starburst galaxies.

\subsection{NGC\,625 As A Wolf-Rayet Galaxy}
\label{S9.2}

The spectrum of the dominant \HII\ region in NGC\,625 published by
{SCM03b}\nocite{skillman03b} showed a broad \HeII\ $\lambda$4686 feature with 
an equivalent width of 5.5 \AA .  According to the criteria of \citet{conti91},
this qualifies NGC\,625 as a Wolf-Rayet (W-R) galaxy.  As 
{SCM03b}\nocite{skillman03b} presented the first high S/N spectra of the major
starburst region, it is not surprising that recent W-R galaxy catalogs (e.g., 
{Schaerer, Contini, \& Pindao 1999b}\nocite{schaerer99b}) have excluded this 
system.  In the absence of a detailed study of the photometry of its stars, one
could infer several properties of the current burst of star formation from its 
optical emission line spectrum.  For example, \citet{schaerer99a} find that the
bursts of star formation in all W-R galaxies have ages between 3 and 6 Myr with
durations in the range 2$-$4 Myr.  Thus, simply the presence of the broad 
\HeII\ $\lambda$4686 feature in the spectrum implies that the starburst in 
NGC\,625 is recent (age $\le$ 6 Myr) and short lived ($\le$ 4 Myr).  While this
may be true of the largest \HII\ region in NGC\,625, to infer that this is 
characteristic of the recent global star formation in NGC\,625 is at odds with 
our observations. 

Is this true of other well-studied nearby starbursting dwarfs?  In several such
systems, there appears to be a discrepancy between the ages of the stellar 
clusters and the ages of the field stars.  In NGC\,1569, \citet{hunter00} find 
ages of most of the young star clusters to be less than 30 Myr (the two 
super-star clusters have ages of $\le$ 7 Myr and 10$-$20 Myr), while 
\citet{greggio98} find that the period of enhanced star formation has lasted 
for $\ge$ 100 Myr.  In NGC\,5253, the ages of the central star clusters range 
from 1 to 8 Myr, while the central ``field'' stars are consistent with ages up 
to 50 Myr \citep{tremonti01}.  {Tremonti \etal}\nocite{tremonti01} suggest that
either the field star populations are created without the uppermost IMF (giving
the appearance of an older population), or that the clusters older than 10 Myr
dissolve into the field.  Note that, based on the optical emission line 
spectrum, \citet{schaerer99a} give burst ages for two components of NGC\,5253 
as 3 Myr and 5 Myr. 

In the case of NGC\,625, it appears that star formation has taken place at an 
elevated rate for at least the last 100 Myr and that the sites of the star 
formation have been both concentrated (like the present dominant \HII\ region) 
but also spread about the disk (like the older event in the SE).  The presence 
of broad \HeII\ $\lambda$4686 in the dominant \HII\ region 
{(SCM03b)}\nocite{skillman03b} indicates that the star formation that gave rise
{\it to the present exciting association} must have been recent ($\le$ 3 Myr) 
and short lived (see arguments in {Schaerer \& Vacca 1998}\nocite{schaerer98} 
and references therein). For example, the \hbeta\ equivalent width of 235 \AA\
implies an age of 3 Myr, assuming an instantaneous burst model.  Thus, if 
NGC\,625 were too distant for a resolved study of its recent star formation, it
would be classified as a W-R galaxy, and the star formation in NGC\,625 would 
be characterized entirely by that associated with the dominant \HII\ region.  
The recent star formation histories of NGC\,1569 and NGC\,5253 are both 
consistent with distributed star formation over a period of roughly 100 Myr or 
more interspersed with the formation of a few clusters.  If the bulk of these 
stars are formed in the field or in associations, then one does not require the
solution of dissolving clusters proposed by \citet{tremonti01}.  In sum, the 
estimates of burst durations and ages implied from W-R star features could 
potentially be be quite misleading in characterizing a burst of star formation 
in a dwarf galaxy.

Is the distinction between the age and duration of a global burst of star 
formation in a dwarf galaxy and the age and duration of a single stellar 
association or cluster in a dwarf galaxy important?  We note a number of cases 
where the difference between a burst duration of 50 Myr and 5 Myr should be 
considered.   For example, \citet{krueger95} modeled only 5 Myr duration bursts
when investigating the spectral energy distributions of blue compact galaxies.
Longer duration bursts would tend to ``dampen'' spectral features and lead to 
lower burst parameter strengths.  In cosmological studies of the luminosity 
functions of galaxies, the degree of ``burstiness'' affects the slope of the 
faint end of the luminosity function.  Often, short duration bursts are assumed
in these types of calculations, and the assumption of a 10 Myr duration is 
supported by the reasoning that at longer times the type II SNe produced by the
burst will heat the ISM preventing any further star formation (e.g., {Ferguson 
\& Babul 1998}\nocite{ferguson98}).  Since we are now observing bursts in dwarf
galaxies where star formation has continued at an elevated rate for several 
tens of Myr, it would appear that such ``self-quenching'' is not necessarily 
an universal property of starbursts in dwarf galaxies.  Although many dwarf 
galaxies may truly have short duration bursts, the distribution of burst 
strengths and durations in dwarf galaxies should still be considered an open 
question.

\section{Conclusions}
\label{S10}

We have presented new HST/WFPC2 imaging of the nearby dwarf starburst galaxy 
NGC\,625.  V and I images are used to model the recent evolution of this 
actively star-forming galaxy.  Our single-star photometry, 80\%\ complete to 
magnitudes of 26.0 (V) and 25.3 (I), extracts information on the relatively 
luminous stars in this nearby galaxy, and we compare these results to stellar 
evolution models and simulations to extract quantitative information on the 
recent star formation history.  The well-defined tip of the red giant branch 
feature in the I-band luminosity function allows us to derive an improved 
distance to NGC\,625 of 3.89$\pm$0.22 Mpc, placing it at the distant end of 
the three-dimensional structure of the Sculptor Group.  We use the spatially 
resolved nature of our photometry to ascertain properties of the star 
formation in NGC\,625 over the past 100 Myr.

There exists a clear and well-defined stellar population gradient in NGC\,625.
The young, MS stars are tightly confined to the main disk and active star 
formation regions.  The slightly older BHeB stars are somewhat more diffuse 
than the MS population, and show an interesting extension to the southeast of 
the main disk.  The oldest RGB stars show a relatively uniform, smooth 
distribution around and within the disk, and are found to the largest 
galactocentric radii probed in these observations.  We note that NGC\,625 is 
embedded in quite a large \HI\ distribution, which continues much further from
the disk (3.4\arcmin, or $\sim$ 6 optical scalelengths; 
{MMHS}\nocite{marlowe97}) than does the stellar population studied here 
{(CCF00)}\nocite{cote00}.  The smoothly distributed red giant population 
resembles that found in the nearby BCD galaxy VII\,Zw\,403.

There is a strong spatial correlation of \halpha\ emission and young, MS stars.
The four detected star formation complexes are correlated with 
MS stars in all cases.  We note that our comparative insensitivity 
to low surface brightness \HII\ regions may bias such a comparison to only the 
least-reddened stars.  We have undertaken radio continuum observations of this
system to further probe the nature of potentially embedded star formation 
complexes. 

We have created a new modeling technique to study the effects of differential 
extinction in resolved stellar populations.  We use this model, with fine 
temporal resolution (10$^5$ years), to create our selection regions for the 
BHeB stars.  Furthermore, we self-consistently calculate the percentage 
contamination of the BHeB sequence by heavily reddened MS stars.  Finally, we
use this modeling to demonstrate the lack of BHeB stars produced by star 
formation events of short ($<$ 10 Myr) duration.  While the generalization of 
this method will be developed further in a future publication, we note that 
the current results are in agreement with the extended burst scenario derived
using the BHeB stars.

A basic model of the recent star formation in NGC\,625, using only the BHeB 
stars, suggests an elevated but declining star formation rate over the 
last 100 Myr.  While we reserve a detailed treatment of the recent star 
formation history for a multicolor dataset, it is clear from these data alone 
that star formation in NGC\,625 was stronger in the recent past ($\sim$ 50 $-$ 
100 Myr ago) than it is in the current epoch.  However, the presence of broad 
$\lambda$\,4686\,\AA\ emission in the spectra of {SCM03b}\nocite{skillman03b} 
suggests vigorous star formation relatively recently ($<$ 6 Myr).  Taken 
together, these points suggest that the use of W-R features to characterize 
burst ages can be somewhat misleading and should be treated carefully when a 
color magnitude diagram study of the resolved stellar populations is not 
available; this point is especially important for galaxies at higher redshift.

The triggering mechanism of the current starburst is not known.  We have 
discussed various scenarios by which the current star formation episode may
have begun.  While the neutral and ionized gas kinematics certainly suggest a 
recent merger, no (luminous) companion galaxy is seen near NGC\,625.  The 
stellar and \HI\ distributions seem to suggest a warp in the main disk, 
perhaps a result of the recent accretion of a lower-mass galaxy or \HI\ cloud.
If the current burst has not been triggered by a minor tidal interaction, we 
may be witnessing the in situ disruption of the main disk of NGC\,625 by the 
current star formation episode.

The current major star formation complex, denoted 
NGC\,625\,A in Figure~\ref{figcap3}(a) and Table~\ref{t4}, is not coincident 
with the \HI\ column density maximum, but rather is slightly displaced toward 
the end of the stellar and \HI\ disk and lies in an area of lower \HI\ column 
density.  Furthermore, we find young stars spread out over the entire disk, 
and, in some cases, quite far from the disk.  These factors suggest that star 
formation may have been widespread throughout NGC\,625 in the recent past; 
our simple model of the star formation suggests elevated, widespread star 
formation over the last 100 Myr at the least. 

Interestingly, NGC\,625 displays a high value of (N/O), comparable to values 
seen in BCD galaxies {(SCM03b)}\nocite{skillman03b}.  Those authors 
interpret this to signify a relatively quiescent (or inefficient) period of 
star formation prior to the current burst, allowing intermediate-mass stars to
enrich the ISM in N compared to the faster-timescale O enrichment expected 
from high-mass stars.  Further investigation of the relation 
between chemical composition and star formation should shed light on the 
universality of such quiescent periods leading to N enhancement. 

Finally, comparing NGC\,625 to other nearby dwarf starbursts shows comparable
star formation scenarios over the last 100 Myr.  In both NGC\,1569 and 
NGC\,5253, star formation appears to have been widespread over the last 100 
Myr, with the formation of clusters perhaps signifying concentrated star 
formation both temporally and spatially.  A similar simple model emerges for 
NGC\,625; we find elevated but declining star formation for the last 100 Myr.  
However, the spectral properties of the major \HII\ region, showing
pronounced W-R features, suggest a short duration for the current burst. This 
discrepancy, in the case of NGC\,625, argues for caution in the interpretation
of the presence of W-R features as a definitive property of a short-duration 
starburst episode.  Taken together, these lines of evidence argue for the 
importance of deep, multicolor datasets that allow the derivation of 
line-of-sight reddening corrections, and therefore accurate spatially resolved 
star formation histories, of these intriguing dwarf starburst systems.

\acknowledgements
The authors appreciate the thorough comments of an anonymous referee that 
helped to improve this work, and thank Abi Saha for beneficial conversations 
that helped to facilitate the data analysis.  Support for this work was 
provided by NASA through grant number GO-8708 from the Space Telescope 
Science Institute, which is operated by AURA, Inc., under NASA contract 
NAS5-26555.  We acknowledge support from NASA through LTSARP grant NAG5-9221.
J.\,M.\,C. is supported by NASA Graduate Student Researchers Program (GSRP) 
Fellowship NGT 5-50346, and is grateful for the hospitality of the Institute 
of Astronomy of Cambridge University and the Ruhr-Universit{\" a}t Bochum, 
where parts of this work were completed.  
B.\,W.\,M. is supported by the Gemini Observatory, which is operated by the 
Association of Universities for Research in Astronomy, Inc., on behalf of the 
international Gemini partnership of Argentina, Australia, Brazil, Canada, 
Chile, the United Kingdom, and the United States of America.
E.\,D.\,S is grateful for the hospitality of the Institute of Astronomy of 
Cambridge University during his sabbatical visit.  This research has made use 
of: the NASA/IPAC Extragalactic Database (NED) which is operated by the Jet 
Propulsion Laboratory, California Institute of Technology, under contract with
the National Aeronautics and Space Administration; NASA's Astrophysics Data 
System;  the SIMBAD database, operated at CDS, Strasbourg, France;  and the 
NASA/IPAC Infrared Science Archive, which is operated by the Jet Propulsion 
Laboratory, California Institute of Technology, under contract with the 
National Aeronautics and Space Administration.



\clearpage
\begin{figure}
\begin{center}
\includegraphics[width=15.0 cm]{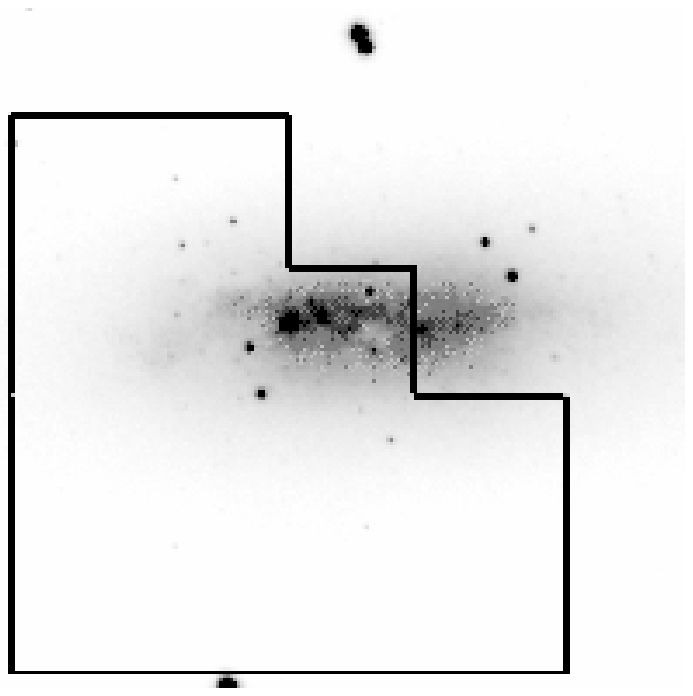}
\end{center}
\caption{Overlay of the HST/WFPC2 field of view on a 4\arcmin$\times$4\arcmin\
R-band image of NGC\,625, taken with the EFOSC2 camera at the ESO 3.6\,m 
telescope (October, 2000; P.I. Bomans); North is up and East is to the left.  
The large star formation region is near the eastern edge of the disk.  Central 
dust concentrations are easily identified, even in this red ground-based image.  
The implications of this are discussed further in the text. Our chosen 
orientation and position allows us to sample the starburst region at high 
resolution (PC chip) while also maximizing our sensitivity to RGB stars away 
from the disk, aiding in our calculation of the galaxy distance in 
\S~\ref{S4}.}
\label{figcap1}
\end{figure}

\clearpage
\begin{figure}
\begin{center}
\includegraphics[width=17.0 cm]{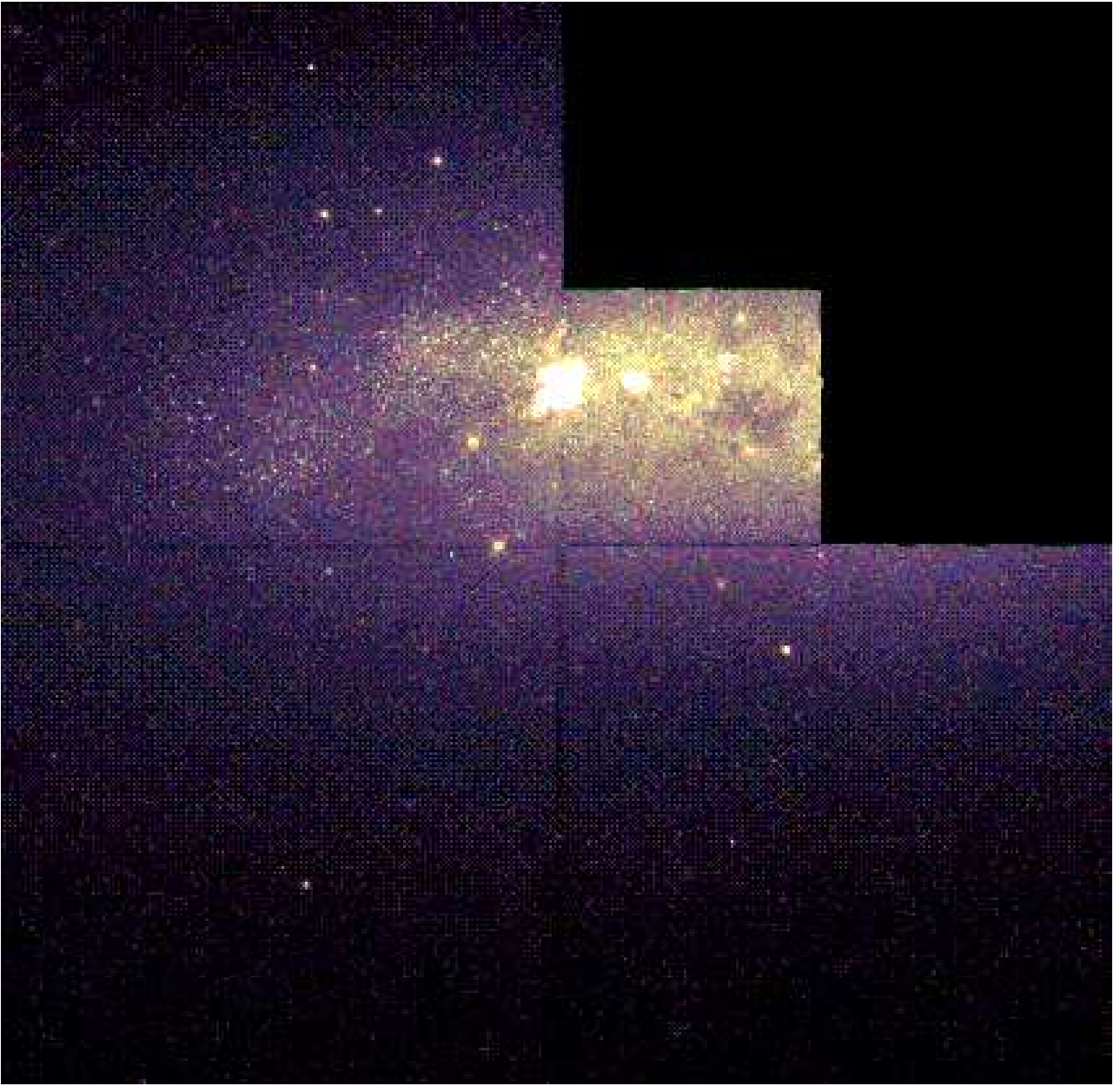}
\end{center}
\caption{4-color image of NGC\,625, created using F555W as blue, 
(F555W $+$ F814W)/2 as green, F814W as red, and F656N as orange.
Note the extremely high stellar density 
throughout the galaxy, and the presence of nebular emission and dust 
obscuration in the central regions.}
\label{figcap2}
\end{figure}

\clearpage
\begin{figure}
\includegraphics[width=17.0 cm]{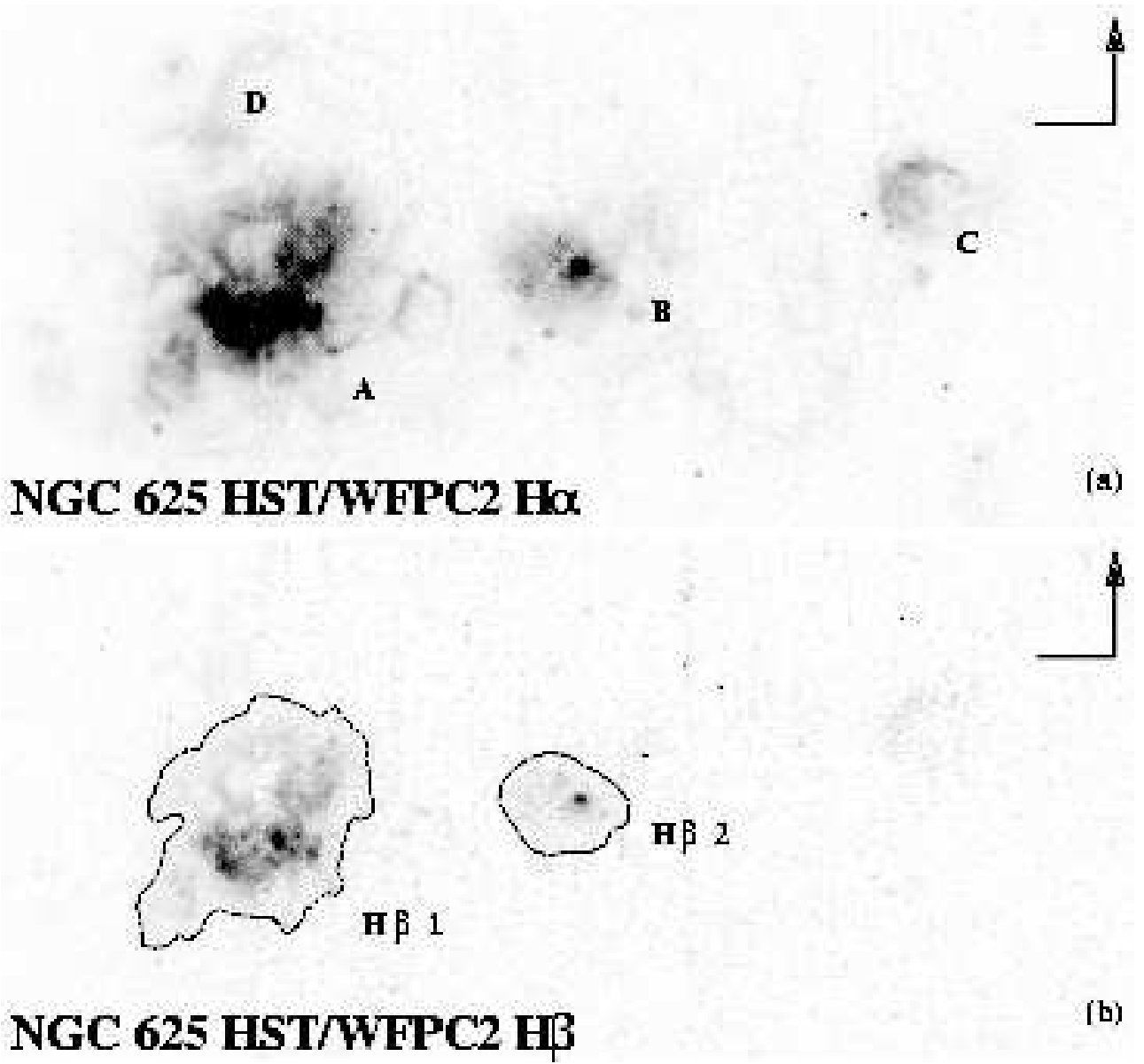}
\caption{Continuum-subtracted images of the central sections of NGC\,625, in 
the lights of \halpha\ (a) and \hbeta\ (b).  The field of view, 39\arcsec\ 
$\times$ 17\arcsec\ (= 736 $\times$ 321 pc), is identical in both images; the 
arrows denote North (tipped) and East.  The linear intensity scale ranges 
from 0 (white) to $\ge$ 8$\times$10$^{-16}$ erg sec$^{-1}$ cm$^{-2}$ (black), 
and is the same in both images.  The labels A-D in (a) correspond to the \HII\
regions described in Table~\ref{t4}.  The labels \hbeta\,1 and \hbeta\,2 in 
(b) correspond to the areas of highest \hbeta\ equivalent width, listed in 
Table~\ref{t5}.  The lower signal to noise ratio of the \hbeta\ image limits 
which sections of the galaxy may be sampled for the effects of internal 
extinction by using the \hab\ ratio.}
\label{figcap3}
\end{figure}

\clearpage
\begin{figure}
\begin{center}
\includegraphics[width=17.0 cm]{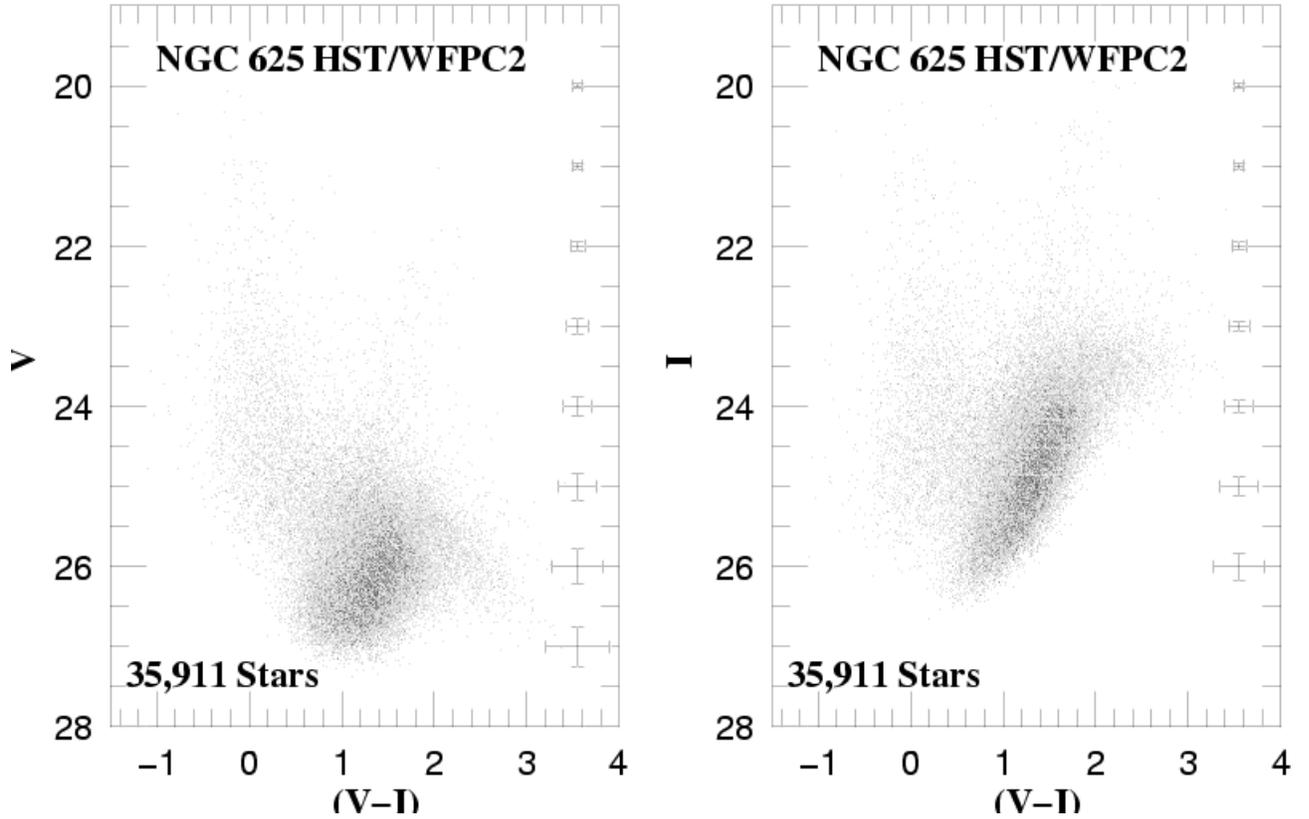}
\end{center}
\caption{Color magnitude diagrams derived from the extracted photometry of 
NGC\,625.  (a) shows the V vs. \vmi\ CMD, and (b) shows the I vs. \vmi\
CMD.  All matched stars with total photometric errors $<$ 0.2 mag in both 
bandpasses are included here (35,911 stars).  Average errors in magnitude
and color (applying bin sizes of 1.0 magnitude) are included at the right 
hand side for comparison.  Salient features of the CMD are easily identified, 
including the blue plume and AGB.  This very red 
AGB (or ``red tail'') extends to \vmi\ $>$ 2.5 and is discussed in 
\S~\ref{S7.2}.}
\label{figcap4}
\end{figure}

\clearpage
\begin{figure}
\begin{center}
\includegraphics[width=12.0 cm]{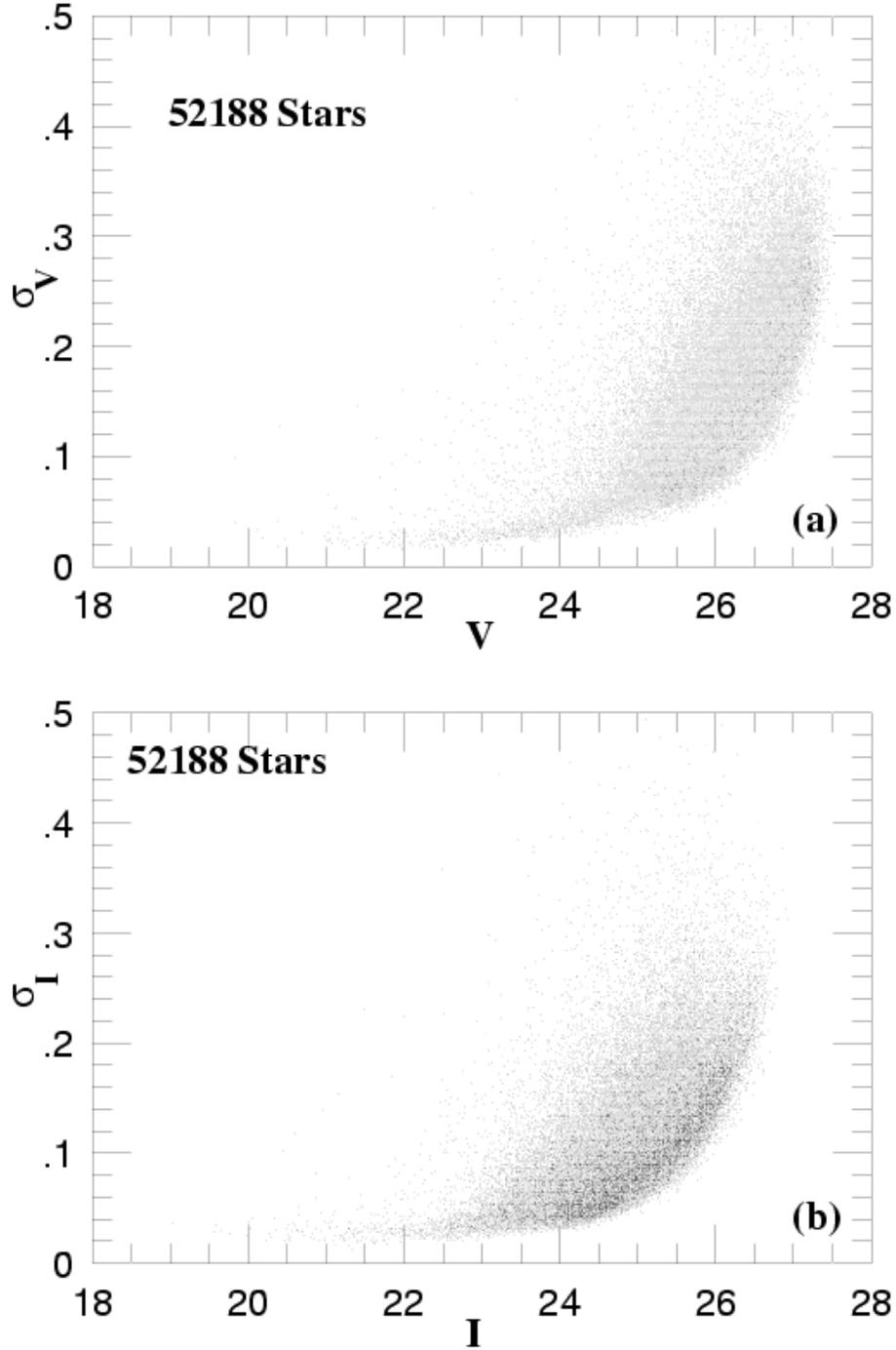}
\end{center}
\caption{DoPHOT internal error distribution for all stars matched between
the V-band (F555W filter, a) and the I-band (F814W filter, b).  
The vertical scatter results from high stellar crowding in some sections 
of the images.  Only stars with photometric errors below 0.2 mag in both 
I and V are kept in the analysis at hand; this reduces the number of
stars to 35911.  See \S~\ref{S3} for further discussion.}
\label{figcap5}
\end{figure}

\clearpage
\begin{figure}
\begin{center}
\includegraphics[width=17.0 cm]{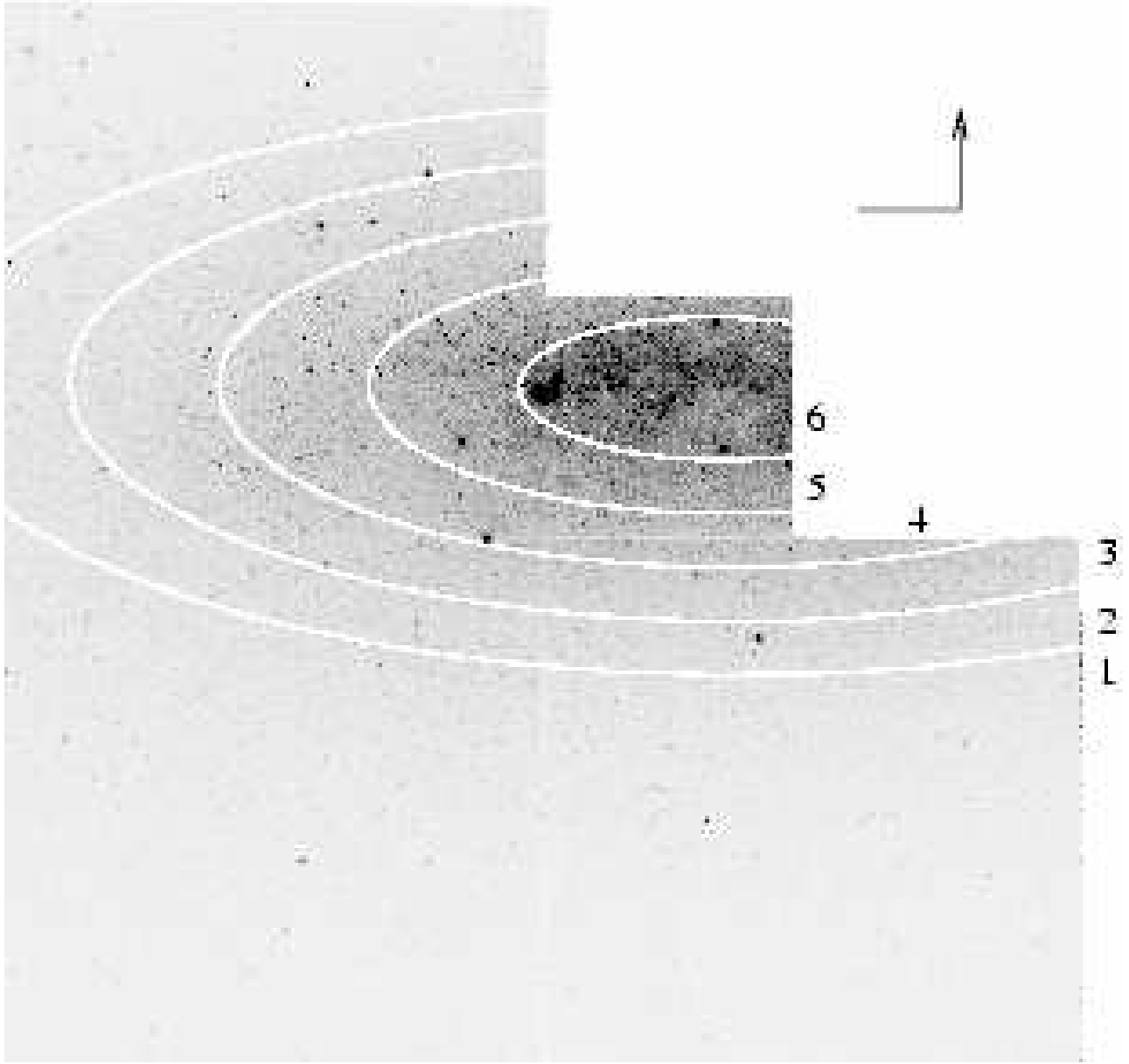}
\end{center}
\caption{NGC\,625 in the I Band (10,400 second exposure); the arrows 
denote North (tipped) and East.  The elliptical contours used in the 
photometric analysis are shown in white; the stellar density increases 
from Region\,1 (essentially halo stars, well separated from the galactic 
disk) to Region\,6 (main disk of the galaxy, including the large \HII\ 
region NGC\,625\,A).  Individual color magnitude diagrams for each of 
these regions are shown in Figure~\ref{figcap7} and discussed throughout
the text.}
\label{figcap6}
\end{figure}

\clearpage
\begin{figure}
\vspace{-0.5 cm}
\begin{center}
\includegraphics[width=10.0 cm]{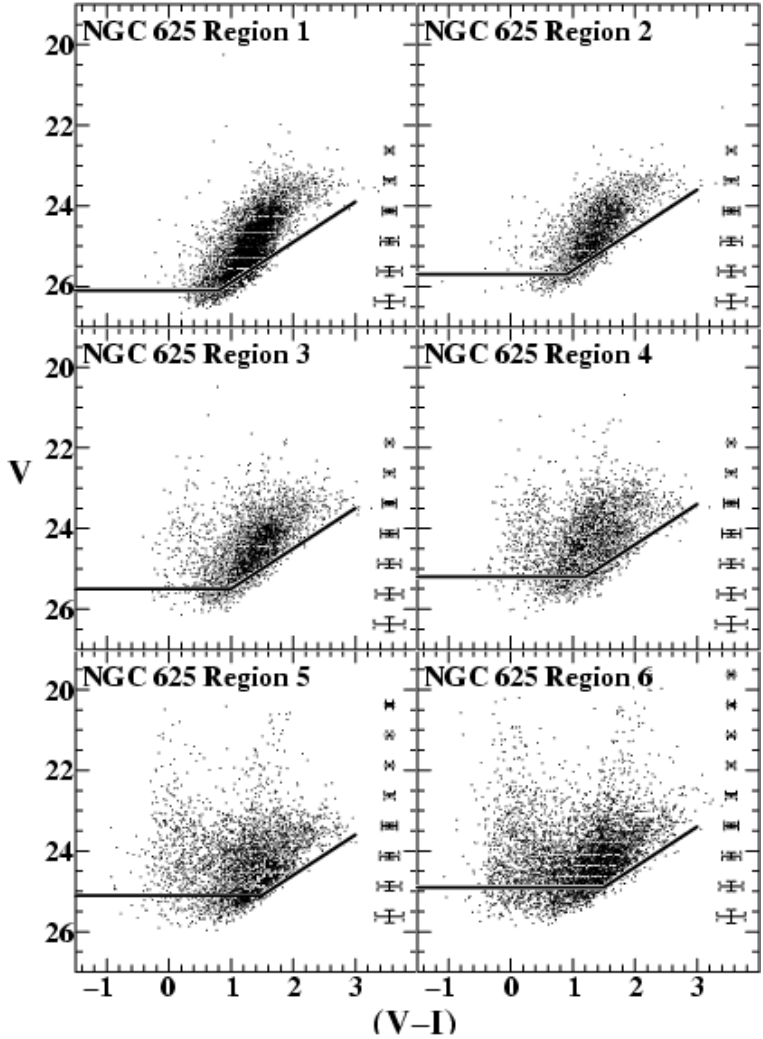}
\end{center}
\caption{CMD's of each of the separate regions of the galaxy, denoted in
Figure~\ref{figcap6}.  The stellar density and crowding are highest in 
Region\,6 (bottom right).  Region\,1, on the other hand, contains mostly halo 
stars (red giants) and is used to calculate the TRGB distance (\S~\ref{S4}) 
and to study the spatial distribution of the old stellar population 
(\S~\ref{S7}). Note the clarity of the TRGB discontinuity at M$_I$ $=$ 23.95 
$\pm$ 0.07 (see Figure~\ref{figcap8}). Only stars with photometric errors $<$ 
0.2 magnitudes in both V and I are displayed here and used for the 
quantitative analysis in deriving the star formation history.  The errorbars 
included at the right in each plot are average photometric and color errors
calculated individually for each region, using magnitude bins of 0.75 
magnitudes.  The thick line in each plot shows the 80\%\ completeness level,
as derived from artificial stars tests.}
\notetoeditor{We request that this figure be printed as a 2-column figure
in the Journal.}
\label{figcap7}
\end{figure}

\clearpage
\begin{figure}
\begin{center}
\includegraphics[width=17.0 cm]{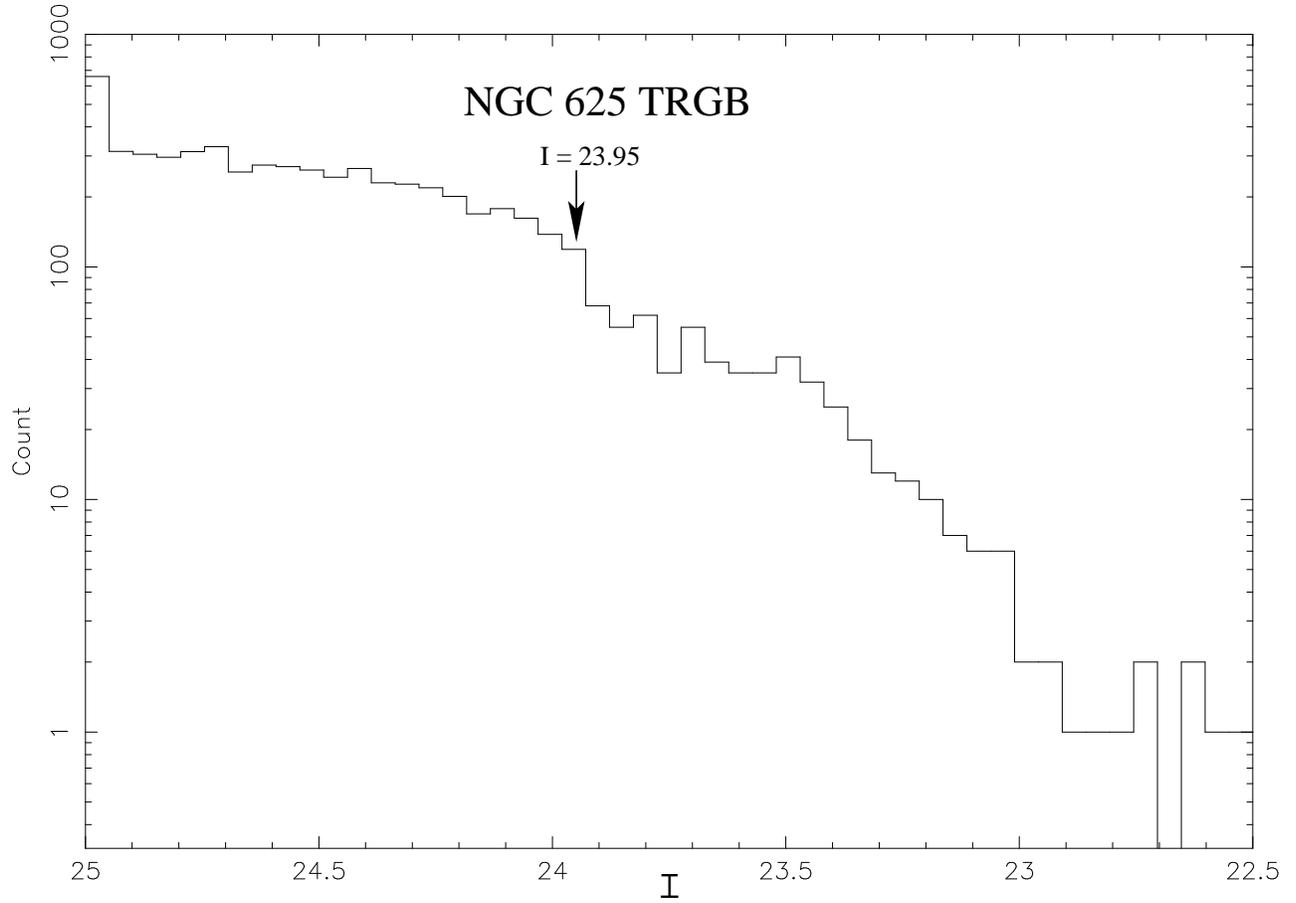}
\end{center}
\caption{I-Band luminosity function of the outer regions of NGC\,625 (Region\,1
shown in Figures~\ref{figcap6} \&\ \ref{figcap7}). The position of the tip of the 
RGB is marked by an arrow at m$_I$ = 23.95$\pm$0.07.  This value is used to 
derive the distance to the galaxy, 3.89$\pm$0.22 Mpc, in \S~\ref{S4}.}
\label{figcap8}
\end{figure}

\clearpage 
\begin{figure}
\begin{center}
\includegraphics[width=17.0 cm]{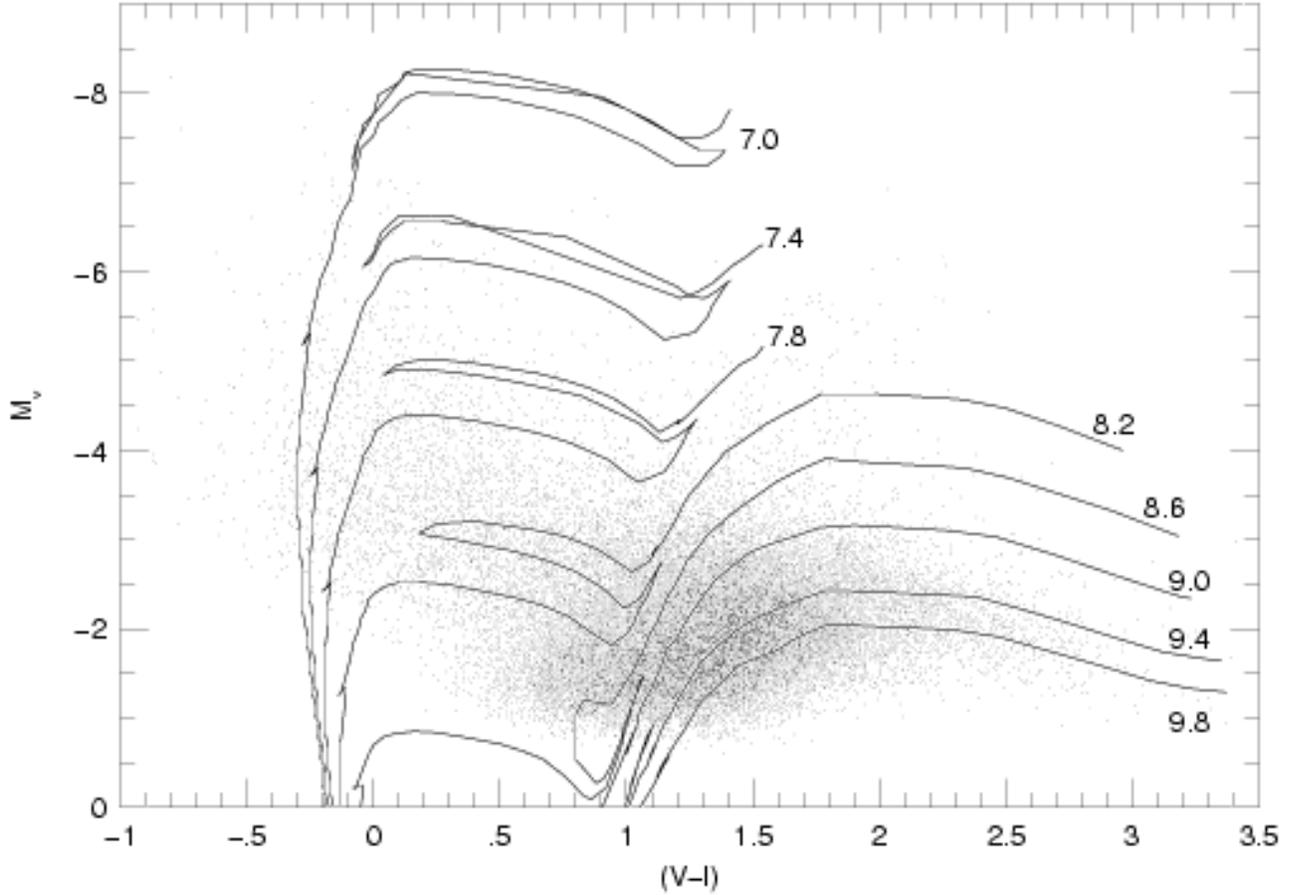}
\end{center}
\caption{Z = 0.004 theoretical isochrones from the Padua group (B94) overlaid 
on the observed V vs. \vmi\ CMD of NGC\,625. Each isochrone has the logarithm 
of its age labeled near the largest value in \vmi.  Note the extended red AGB 
stars detected, and the goodness with which the higher-age isochrones fit this 
feature of the CMD.  Lower-metallicity isochrones do not significantly populate
this region of the CMD.  This suggests the presence of a relatively large 
intermediate age ($\sim$ few Gyrs) population of AGB stars in NGC\,625.}
\notetoeditor{The B94 reference in this Figure caption should be referenced to 
Bertelli \etal (1994).  Using the ``nocite'' command in the figure caption 
causes errors upon compiling the latex file.}
\label{figcap9}
\end{figure}

\clearpage
\thispagestyle{empty}
\begin{figure}
\vspace{-1.0 cm}
\begin{center}
\includegraphics[height=11.5 cm]{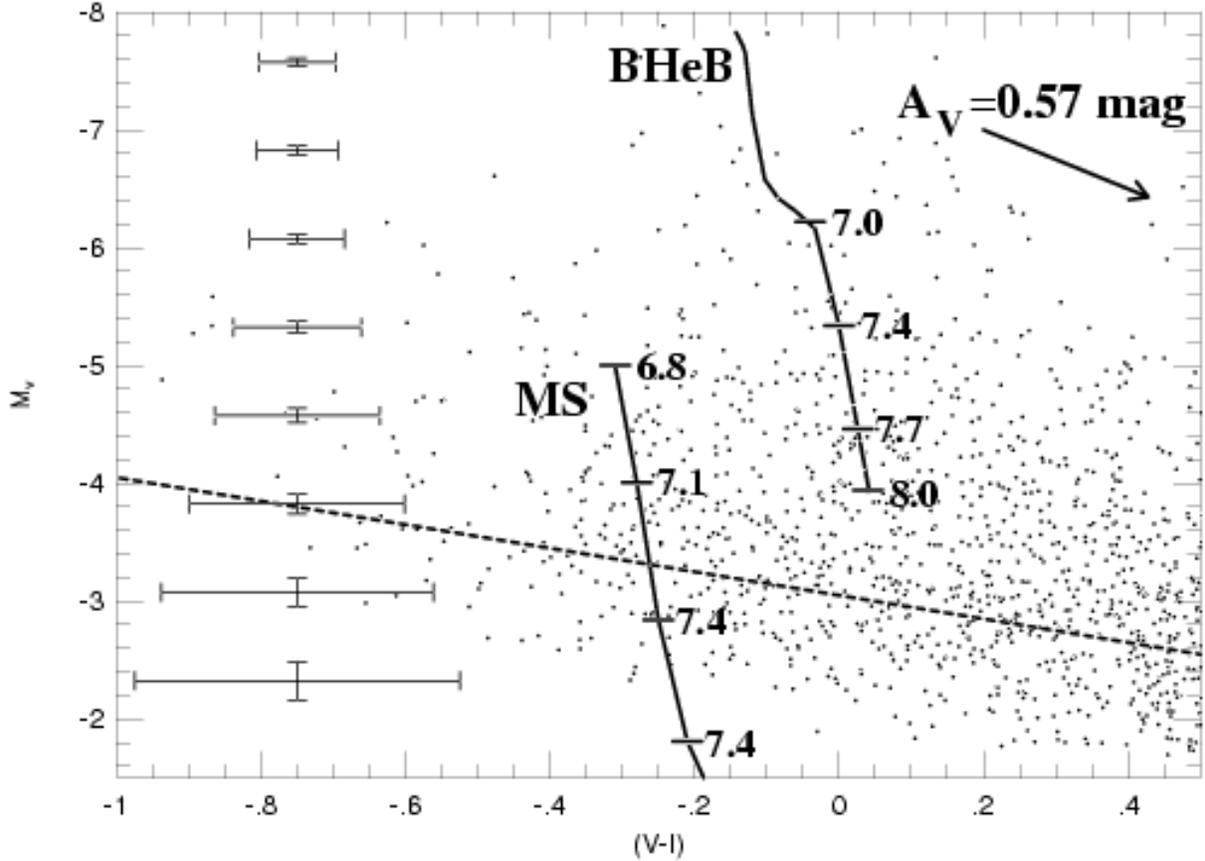}
\end{center}
\vspace{-1.0 cm}
\caption{Closer view of the blue plume region of the V vs. \vmi\ CMD, extracted
from the central star formation regions (Region\,6; see Figure~\ref{figcap6}).
Overlaid are lines denoting the zero-age MS and the position of the 
blue extent of the BHeB; tick marks label logarithmic ages. The dotted line 
shows the 80\%\ completeness level for the Region\,6 photometry, as derived 
from artificial stars tests; note that BHeB stars with ages $<$ 100 Myr are not
appreciably affected by incompleteness.  The wide color spread of the blue 
plume, which should only contain these relatively tight sequences of stars, is 
attributed to large amounts of differential extinction throughout the 
star-forming regions.  The arrow at the upper left indicates the largest 
reddening detected in our narrow-band imaging (see \S~\ref{S3.3}). Since 
regions of low reddening were also found, the differential extinction is large
enough to significantly blur the positions of these stars in the color 
magnitude diagram, forcing us to apply a statistical separation of these 
populations.}
\label{figcap10}
\end{figure}

\clearpage
\begin{figure}
\begin{center}
\includegraphics[width=17.0 cm]{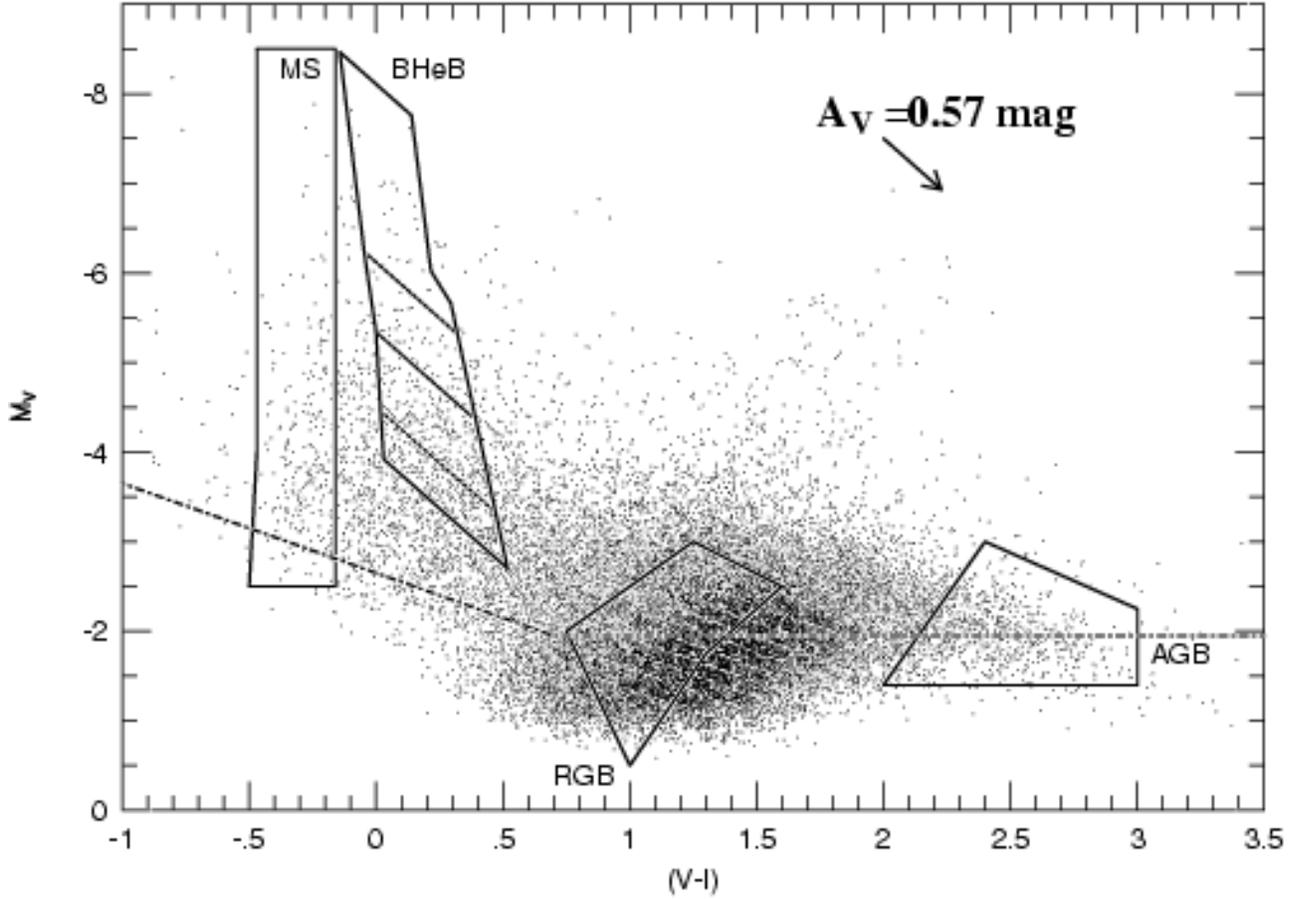}
\end{center}
\caption{The regions of the CMD used to isolate the MS, BHeB, RGB, and AGB
populations.  The MS and BHeB regions are based on the Z=0.004 isochrones of 
B94; the RGB and AGB regions are created by hand, but follow the empirical 
distribution of RGB stars and the theoretical and empirical distributions of 
AGB stars (see discussion in \S~\ref{S5.1}).  The effects of differential 
extinction have been modeled as described in \S~\ref{S5.2}.  The dotted lines
in the BHeB sequence represent the borders between the 0-25, 25-50, 50-75, and 
75-100 Myr age populations, respectively.  Note that these borders follow the 
direction of the reddening vector. The dot-dash line indicates the 80\%\ 
completeness level for the entire galaxy, as derived from artificial star
tests.}
\notetoeditor{The B94 reference in this Figure caption should be referenced to 
Bertelli \etal (1994).  Using the ``nocite'' command in the figure caption 
causes errors upon compiling the latex file.}
\label{figcap11}
\end{figure}

\clearpage
\begin{figure}
\begin{center}
\includegraphics[width=13.0 cm]{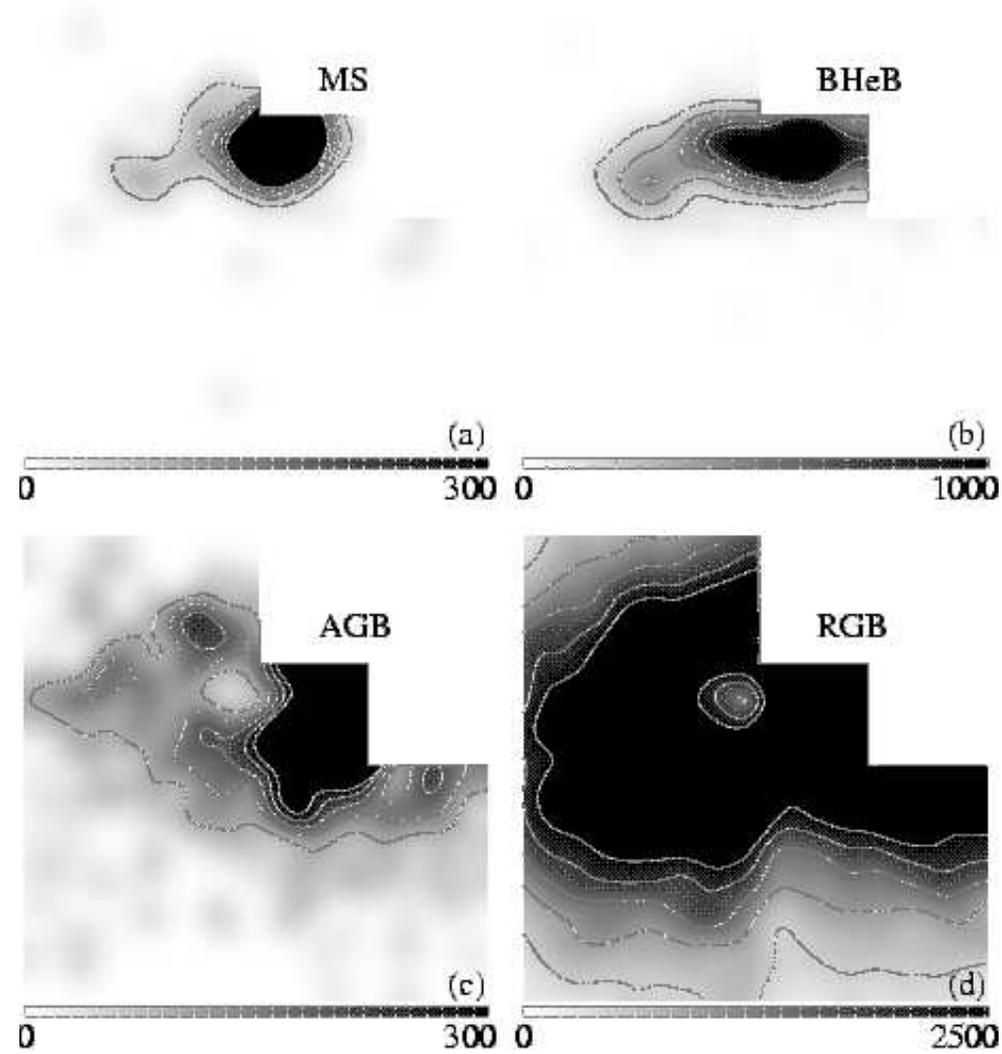}
\end{center}
\caption{Comparison of the spatial distribution of the four different 
populations of stars studied in NGC\,625, extracted using the CMD regions 
shown in Figure~\ref{figcap11}.  These density plots show the number of 
\stpkpc\ for each population, with the color bar below each image denoting
intensity.  (a) shows the MS stars (contours at 50, 100, 150, 200 \&\ 250 
\stpkpc), (b) shows the BHeB stars (contours at 150, 300, 450, 600 \&\ 
750 \stpkpc), (c) shows the AGB stars (contours at 100, 150, 200, 250 \&\ 
300 \stpkpc), and (d) shows the RGB stars (contours at 500, 1000, 1500, 2000 
\&\ 2500 \stpkpc).  The field of view is identical in all images, with north 
up and east to the left.  Note that the distribution of RGB stars is 
relatively uniform and diffuse, while the populations grow more compact as 
one moves toward younger age stars.}
\label{figcap12}
\end{figure}

\clearpage
\begin{figure}
\begin{center}
\includegraphics{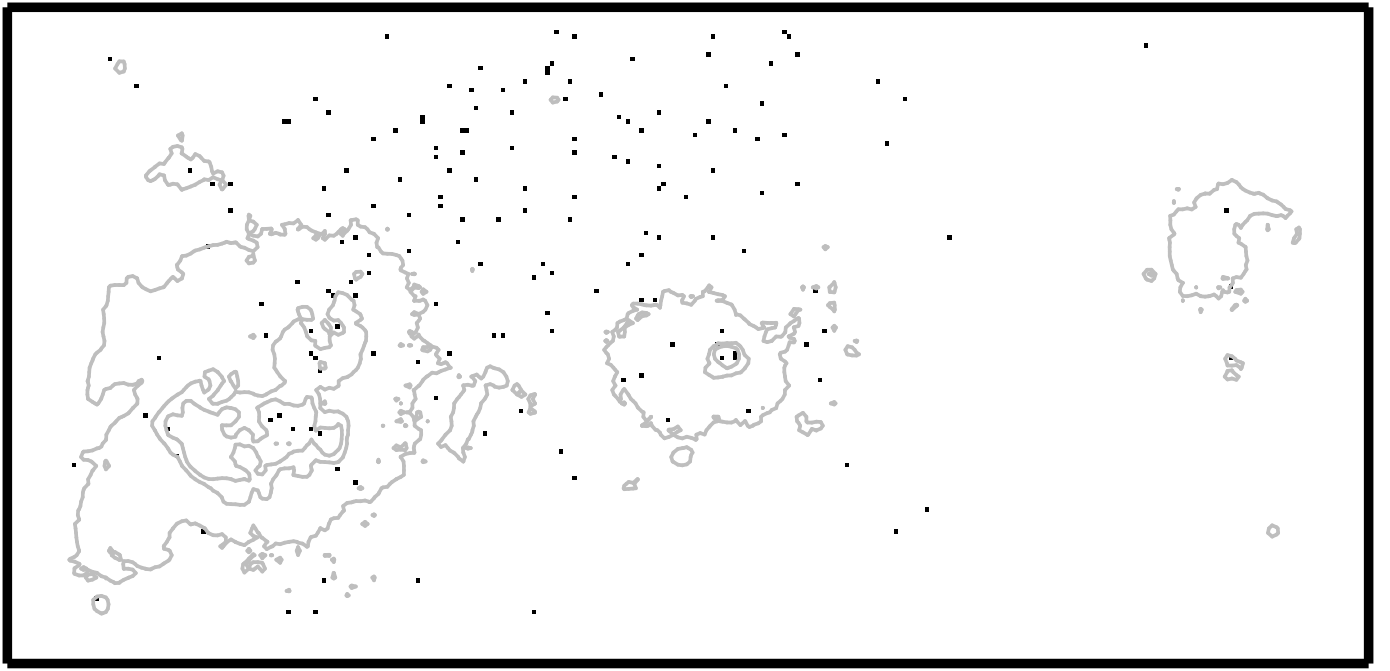}
\end{center}
\caption{Comparison of the spatial distribution of young MS stars 
(M$_V$ $<$ -2.5; black dots) with the morphology of high-surface brightness
\halpha\ emission (gray contours).  The field of view is oriented with North 
up and East to the left, and is approximately 32\arcsec$\times$17\arcsec\ 
(600$\times$320 pc at the distance of 3.89 Mpc derived in \S~\ref{S4}).  The 
three large \HII\ complexes are easily identified with those labeled as A, 
B and C in Figure~\ref{figcap3}(a).  The \halpha\ contours are at levels of 
(1$\times$10$^{-16}$, 5$\times$10$^{-16}$, and 1$\times$10$^{-15}$) 
erg\,sec$^{-1}$\,cm$^{-2}$.  Each \HII\ region is identified with present-day 
MS stars.  The improved surface brightness sensitivity of the ground-based
imaging of SCM03a demonstrates that diffuse \halpha\ 
emission, not detected here, is found throughout (and slightly beyond) this 
field of view.}
\notetoeditor{The SCM03a reference in this Figure caption should be referenced 
to Skillman, Cote \& Miller (2003a).  Using the ``nocite'' command in the 
figure caption causes errors upon compiling the latex file.}
\label{figcap13}
\end{figure}

\clearpage
\thispagestyle{empty}
\begin{figure}
\vspace{-1.0 cm}
\begin{center}
\includegraphics[width=16.0 cm]{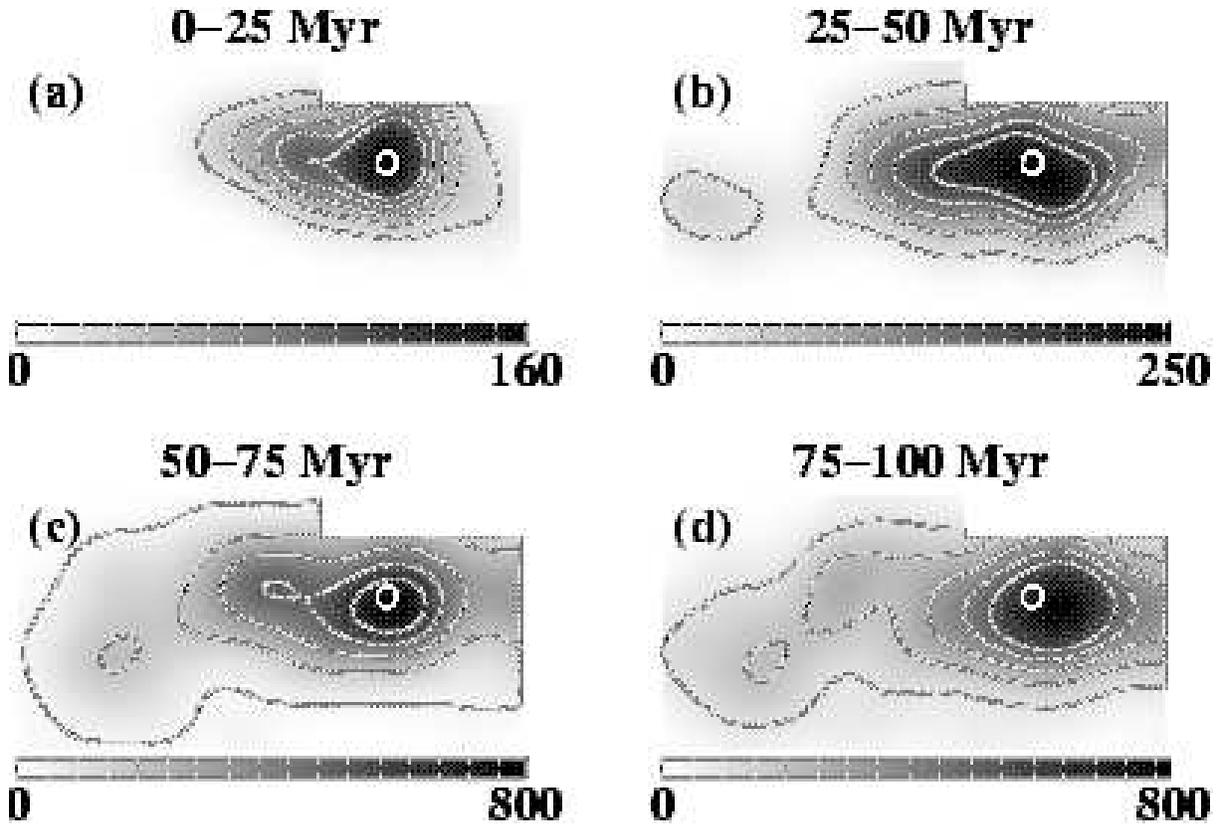}
\end{center}
\caption{Spatial distribution of star formation over the last 100 Myr, as 
derived from the BHeB stars (see \S~\ref{S6.3}).  In each plot, the number 
of BHeB \stpkpc\ is plotted, with intensities indicated by the color bar
under each image; the field of view is $\sim$ 1.7 $\times$ 0.9 kpc, with 
North up and East to the left.  (a) shows the youngest stars, of age 0-25 Myr; 
note the concentrated star formation in the main giant \HII\ region; (b) shows 
stars stars of age 25-50 Myr, and clearly demonstrates that star formation has 
been occurring in different regions of the galaxy during this epoch; (c) and (d)
show older stars, of ages 50-100 Myr, and demonstrate that the star formation
rate has been elevated for this entire 100 Myr period.  The white circle in 
each plot is the highest contour for the young, 0-25 Myr population; comparing 
the location of this peak with past star formation peaks also shows that star
formation has been moving throughout the disk over the last 100 Myr.  The lowest 
contour in each plot demonstrates the extent of low-level star formation 
throughout these epochs.  The contour levels (in number of \stpkpc)
in each image are:  (a) 20, 40, 60, 80, 100; (b) 30, 70, 110, 150, 190; (c) 40,
200, 360, 520, 680; (d) 100, 205, 310, 415, 520.}
\label{figcap14}
\end{figure}

\clearpage
\begin{figure}
\begin{center}
\includegraphics[width=9.0 cm]{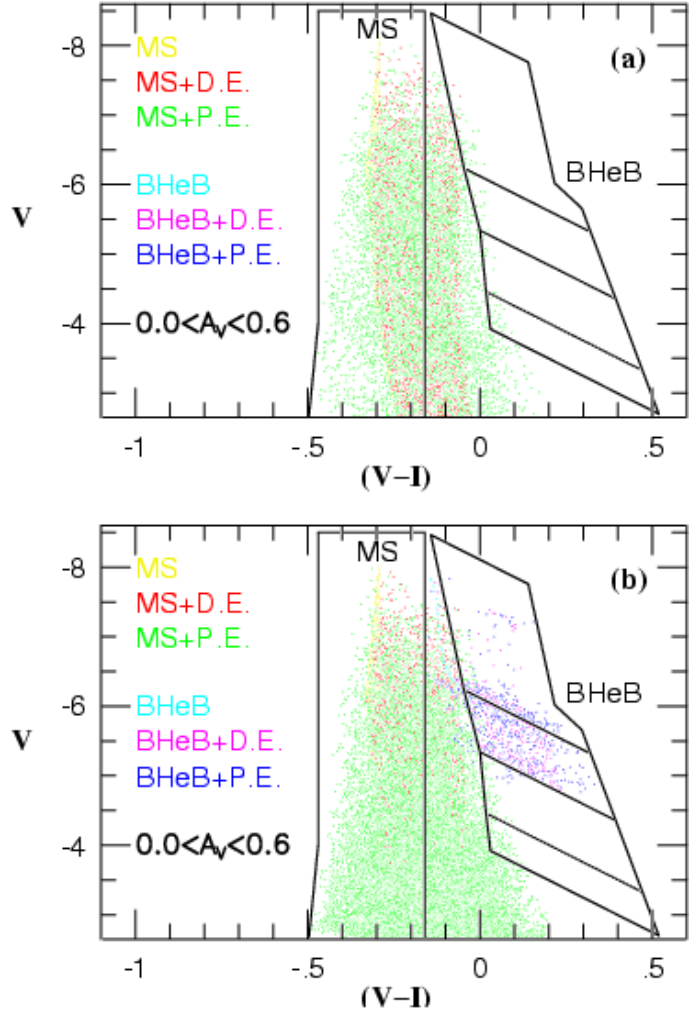}
\end{center}
\caption{Comparison of synthetic bursts of star formation, with equal (constant) 
star formation rates, but different temporal durations.  (a) shows a burst of 
duration 5.0 Myr (i.e., a standard ``short-burst'' duration as is typical in 
models of starburst parameters), while (b) shows a burst of duration 50.0 Myr.
Different colors correspond to different evolutionary stages, as labeled;
``D.E.'' corresponds to differential extinction, and ``P.E.'' corresponds to 
photometric errors.  The MS and BHeB selection regions are shown in bold lines.  
The dotted lines within the BHeB regions correspond to the four different age bins 
(see \S~\ref{S6.3}).  It is clear that, using the Z=0.004 stellar evolution models 
of B94, no substantial numbers of BHeB stars can be produced
unless the star formation event is extended temporally.  Comparing to the many
empirical stars in the BHeB region of the CMD, we conclude that the star 
formation in NGC\,625 has extended over the last 100 Myr.}
\notetoeditor{The B94 reference in this Figure caption should be referenced to 
Bertelli \etal (1994).  Using the ``nocite'' command in the figure caption 
causes errors upon compiling the latex file.}
\label{figcap15}
\end{figure}

\clearpage
\begin{deluxetable}{cccc}
\tabletypesize{\scriptsize}
\tablecaption{Basic Properties of NGC\,625}
\tablewidth{0pt}
\tablehead{
\colhead{Property} 
&\colhead{Value} 
&\colhead{Reference}
&\colhead{Comments}}
\startdata
Mass 			
&1.1$\times$10$^8$ \msun 	&{CCF00}\nocite{cote00}	
&\HI\ mass only; See \S~\ref{S7.1}\\
V$_{Helio}$		&406 \kms			
&{CCF00}\nocite{cote00}	&\\
Distance 		&3.89 Mpc  			
&This work 	&See \S~\ref{S4}\\
M$_B$			&-16.28				
&{MMHS}\nocite{marlowe97} 		&\\
E(B-V)	&0.016				
&{SFD98}\nocite{schlegel98}		
&Foreground only; see \S~\ref{S3.1}\\
12+log(O/H)		&8.14			
&{SCM03b}\nocite{skillman03b}		
&See \S~\ref{S4}\\
\enddata
\tablerefs{CCF00 - \citet{cote00}; MMHS - \citet{marlowe97}; SFD98 - 
\citet{schlegel98}; SCM03b - \citet{skillman03b}}
\label{t1}
\end{deluxetable}

\clearpage
\begin{deluxetable}{ccccc}
\tabletypesize{\scriptsize}
\tablecaption{HST WFPC2 Observations of NGC\,625, Program 
GO-8708\tablenotemark{a}}
\tablewidth{0pt}
\tablehead{
\colhead{Filter} 
&\colhead{Emission} 
&\colhead{Date}   
&\colhead{Exposure (sec)}
&\colhead{Datasets}}
\startdata
F656N 		&\halpha &2000 Sep 23 &800   &U64V0207R, U64V0208R\\
F487N 		&\hbeta  &2000 Sep 23 &1200  &U64V0205R, U64V0206R\\ 
F555W 		&V-Band  &2000 Sep 23 &5200  &U64V0201R - U64V0204R\\ 
F814W 		&I-Band  &2000 Sep 24 &10400 &U64V0101R - U64V0108R\\ 
\enddata
\tablenotetext{a}{When comparing astrometry values derived from
these images with other data, we note a 5.1\arcsec\ offset in declination.
The cause of this error is unknown, but the astrometry derived in this paper
has been drawn from a coordinate solution using separate imaging.  Further 
use of these data for astrometric purposes should note this difficulty.}
\label{t2}
\end{deluxetable}

\clearpage
\begin{deluxetable}{lccccccc}
\tabletypesize{\scriptsize}
\tablecaption{DoPHOT PSF Shape Parameters and Aperture Corrections}
\tablewidth{0pt}
\tablehead{
\colhead{Chip} &\colhead{Filter} &\colhead{FWHM}   
&\colhead{$\beta_4$} &\colhead{$\beta_6$} &\colhead{$\beta_8$} 
&\colhead{N\tablenotemark{a}} &\colhead{Residual}\\
& &\colhead{(pixels)} & & & &{(Stars)}	&{(rms)}}
\startdata
PC 		&F555W  &2.0 &7.6 &-8.3 &3.9 &63   &0.296\\
PC 		&F814W  &2.6 &7.6 &-8.3 &3.9 &141  &0.196\\
WF2 		&F555W  &1.7 &3.6 &-1.1 &0.4 &93   &0.162\\ 
WF2 		&F814W  &1.9 &3.6 &-1.1 &0.4 &274  &0.125\\ 
WF3 		&F555W  &1.9 &3.6 &-1.1 &0.4 &131  &0.147\\ 
WF3		&F814W  &2.1 &3.6 &-1.1 &0.4 &326  &0.122\\
WF4 		&F555W  &1.8 &3.6 &-1.1 &0.4 &107  &0.142\\ 
WF4 		&F814W  &2.0 &3.6 &-1.1 &0.4 &343  &0.132\\
\enddata
\tablenotetext{a}{Number of isolated stars which were used to 
determine the aperture correction.  See further discussion in
\S\ref{S3.1}.}
\label{t3}
\end{deluxetable}

\clearpage
\begin{deluxetable}{lccccccc}
\tabletypesize{\scriptsize}
\tablecaption{The Bright \HII\ Regions of NGC\,625}
\tablewidth{0pt}
\tablehead{\colhead{Feature\tablenotemark{a}}        
&\colhead{Center RA}     
&\colhead{Center DEC}  
&\colhead{Diameter\tablenotemark{b}}
&\colhead{Diameter\tablenotemark{c}}
&\colhead{\halpha\ Flux} 
&\colhead{L$_{\rm H\alpha}$\tablenotemark{c}}
&\colhead{SFR\tablenotemark{d}}\\
&\colhead{(J2000)} 
&\colhead{(J2000)} 
&\colhead{(arcsec)} 
&\colhead{(pc)}
&\colhead{(10$^{-14}$\,erg\,sec$^{-1}$\,cm$^{-2}$)}
&\colhead{(10$^{36}$\,erg\,sec$^{-1}$)}
&\colhead{(\msun\,yr$^{-1}$)}}
\startdata
NGC\,625\,A 	&1:35:06.8	&-41:26:13.0	&7.2 &136	
&204$\pm$10     &3700		& \nodata\\
NGC\,625\,B 	&1:35:06.0	&-41:26:11.2	&3.3 &62	
&112$\pm$6 	&2000		& \nodata\\
NGC\,625\,C 	&1:35:04.9	&-41:26:09.3	&1.7 &32	
&7.8$\pm$0.4 	&140 		& \nodata\\
NGC\,625\,D 	&1:35:07.0	&-41:26:05.9	&1.2 &23	
&6.6$\pm$0.3 	&120		& \nodata\\
Total Galaxy    & \nodata         & \nodata     & \nodata 
& \nodata   &350$\pm$18     &6300		&5.0$\times$10$^{-2}$\\
\enddata
\tablenotetext{a}{See Figure~\ref{figcap3} for locations.  Note that 
these \HII\ regions (A, B, C, D) correspond, roughly, to regions 5, 9, 
18, and 4 in {SCM03a}\nocite{skillman03a}.  The larger number of \HII\ 
regions in that study is a result of higher sensitivity to low surface 
brightness features.}
\tablenotetext{b}{Estimated assuming a circular morphology for the
\HII\ region; note that, in Figure~\ref{figcap3}, the apertures used to 
calculate the flux from each region follow the contours of high 
equivalent width and are not necessarily circular.}
\tablenotetext{c}{Calculated for the TRGB distance derived in 
\S~\ref{S4} of 3.89$\pm$0.22 Mpc.}
\tablenotetext{d}{Calculated from the relation SFR = 
(7.94$\times$10$^{-42}$ erg$^{-1}$ sec)$\cdot$L$_{\rm H\alpha}$,
in units of \msun\ yr$^{-1}$; see {Kennicutt, Tamblyn, \& Congdon 
(1994)}\nocite{kennicutt94}.}
\label{t4}
\end{deluxetable}

\clearpage
\begin{deluxetable}{cccccc}
\tabletypesize{\scriptsize}
\tablecaption{Narrow Band Photometry of NGC\,625}
\tablewidth{0pt}
\tablehead{\colhead{Feature\tablenotemark{a}}        
&\colhead{\halpha\ Flux}     
&\colhead{\hbeta\ Flux}  
&\colhead{EW(\hbeta)\tablenotemark{b}} 
&\colhead{\hab\ Flux}
&\colhead{A$_V$\tablenotemark{d}}\\
&\colhead{(10$^{-14}$\,erg\,sec$^{-1}$\,cm$^{-2}$)} 
&\colhead{(10$^{-14}$\,erg\,sec$^{-1}$\,cm$^{-2}$)} 
&\colhead{(\AA)} 
&\colhead{Ratio\tablenotemark{c}}
&\colhead{(mag)}}
\startdata
NGC\,625 \hbeta\,1 	&180$\pm$9.0  &52.6$\pm$2.6  &420$\pm$21 
&3.43$\pm$0.05 &0.42$\pm$0.03\\
NGC\,625 \hbeta\,2 	&25.3$\pm$1.3 &6.9$\pm$0.3   &217$\pm$11 
&3.64$\pm$0.05 &0.57$\pm$0.03\\
\enddata
\tablenotetext{a}{See Figures~\ref{figcap3}(a) and \ref{figcap3}(b) for 
aperture locations.  Note that these regions are smaller than the \HII\ 
regions characterized in Table~\ref{t4}, and sample only the areas with the 
highest \hbeta\ equivalent widths.  The low signal-to-noise ratio of the
\hbeta\ image therefore offers the limiting term in our ability to sample
internal variations in the extinction properties of NGC\,625.}
\tablenotetext{b}{Average value over the aperture, rounded to the nearest 
Angstrom; errors are assumed at the 5\%\ level.}
\tablenotetext{c}{The error terms account for Poisson errors, the 
uncertainty in continuum subtraction, and for potential CTE effects
across the individual features considered.  For a more comprehensive
discussion of the individual error terms, see \S~\ref{S3}.}
\tablenotetext{d}{Corrected for 0.05 mag of foreground extinction 
{(SFD98)}\nocite{schlegel98}.}
\label{t5}
\end{deluxetable}

\clearpage
\begin{deluxetable}{cccc}
\tabletypesize{\scriptsize}
\tablecaption{RGB:AGB Number Ratios In Photometric Regions}
\tablewidth{0pt}
\tablehead{
\colhead{Region\tablenotemark{a}} 
&\colhead{$\sigma$(V,I)\tablenotemark{b}\,\,\, $<$ 0.10} 
&\colhead{$\sigma$(V,I) $<$ 0.15}   
&\colhead{$\sigma$(V,I) $<$ 0.20}}
\startdata
Region\,1 		&39.8$\pm$6.5   &40.1$\pm$4.2   &51.1$\pm$5.0\\
Region\,2 		&22.2$\pm$4.5   &25.9$\pm$3.3   &29.3$\pm$3.1\\ 
Region\,3 		&31.9$\pm$8.1   &21.0$\pm$2.4   &22.5$\pm$2.1\\ 
Region\,4 		&41.8$\pm$18.9  &19.3$\pm$2.7   &16.1$\pm$1.6\\
Region\,5 		&7.4$\pm$1.8    &8.3$\pm$0.9    &8.9$\pm$0.8\\ 
Region\,6		&8.2$\pm$2.2    &7.4$\pm$0.8    &1.5$\pm$0.1\\
\enddata
\tablenotetext{a}{See Figure~\ref{figcap6} for the location of each region.}
\tablenotetext{b}{The different values of photometric errors in each column
correspond to different maximum values of errors in V or I which a given star
was allowed to have, and remain in the calculation.  So, for example, 
$\sigma$(V,I) $<$ 0.10 means that all stars have total photometric errors 
less than 0.1 magnitudes in both V and I.  Note that small number statistics 
begin to affect the error terms for the lowest photometric error cut (e.g.,
Regions\,3, 4).}
\label{t6}
\end{deluxetable}

\clearpage
\thispagestyle{empty}
\rotate
\begin{deluxetable}{cccccccccc}
\tabletypesize{\tiny}
\tablecaption{Property Comparison of Selected Nearby Starburst Galaxies}
\tablewidth{0pt}
\tablehead{
\colhead{Property\tablenotemark{a}} 
&\colhead{NGC\,625} 
&\colhead{NGC\,1569}
&\colhead{NGC\,1705}
&\colhead{UGC\,4483}
&\colhead{M\,82}
&\colhead{VII\,Zw\,403}
&\colhead{NGC\,4214}
&\colhead{NGC\,4449}
&\colhead{NGC\,5253}}
\startdata
\HI\ Mass (\msun)		
&1.1$\times$10$^8$\tablenotemark{b}\,\, (1) 	
&1.3$\times$10$^8$ (2)
&1.5$\times$10$^8$ (3)	
&3.7$\times$10$^7$ (4) 	
&7.8$\times$10$^8$ (5)	
&3.2$\times$10$^7$\tablenotemark{c}\,\, (6) 	
&1.0$\times$10$^9$ (7) 	
&2.3$\times$10$^9$$\times$h$^2$ (8) 	
&$\frac{3{\times}10^8}{{sin^2}(i)}$ (9)\\

V$_{Helio}$ (\kms)	
&406 (1)	
&$-$104 (10)	
&628 (11)	
&178 (12)
&203 (13)	
&$-$100 (14) 	
&291 (13)  	
&207 (10) 	
&404 (13)\\

Distance (Mpc)		
&3.89$\pm$0.22 (15)  	
&2.2 $\pm$0.6 (16)	
&5.1$\pm$0.6 (17) 
&3.4$\pm$0.2 (18)  	
&3.9$\pm$0.4 (19)
&4.34$\pm$0.07 (20)  	
&2.7$\pm$0.3 (21)  	
&2.9\tablenotemark{d}\,\,\,(22)  	
&4.1$\pm$0.2 (23)\\

M$_B$\tablenotemark{e}			
&$-$16.3 (24)		
&$-$14.9 	
&$-$16.3 (24)	
&$-$12.5	
&$-$18.7 	
&$-$13.7   	
&$-$16.9	
&$-$17.3	
&$-$17.1 (24)\\

Gal. Lat. (\degree)	
&$-$73.1		
&11.2	
&$-$38.7 
&34.9
&40.7	
&37.3	
&78.1	
&72.4	
&30.1\\

E(B-V)\tablenotemark{f}		
&0.016 			
&0.695  
&0.008  
&0.034 	
&0.156
&0.037 	
&0.022 	
&0.019  
&0.074\\

12+log(O/H)		
&8.14$\pm$0.02 (25)	
&8.19$\pm$0.02 (26)	
&8.0\tablenotemark{d}\,\, (27) 	
&7.52$\pm$0.03 (28)	
&$\gsim$ 8.9 (29) 	
&7.69$\pm$0.01 (30)   	
&8.29$\pm$0.03 (31)
&8.32\tablenotemark{d}\,\, (32)	
&8.15$\pm$0.04 (33)\\

Current SFR (\msun\ yr$^{-1}$)
&0.05 (15)		
&0.1\tablenotemark{g}\,\, (34)	
&0.1\tablenotemark{h}\,\, (24)	
&0.001\tablenotemark{i}\,\, (28)	
&0.7\tablenotemark{i}\,\, (35)	
&0.01\tablenotemark{g}\,\, (34)     	
&0.1\tablenotemark{g}\,\, (34)	
&0.2\tablenotemark{g}\,\, (34)	
&0.2\tablenotemark{g}\,\, (34)\\

100 $\mu$m IRAS Luminosity\\
(10$^{36}$ erg/sec)\tablenotemark{j}
&140
&270
&45
&Not\,detected
&19000
&Not\,detected
&200
&Not\,detected
&530\\

X-Ray Luminosity\\
(10$^{36}$ erg/sec)
&100 (36)
&800 (37)
&120 (38)
&N/A
&40000 (39)
&190 (40)
&180 (41)
&2500 (42)
&650 (43)\\

\enddata
\vspace{-0.5 cm}
\tablerefs{
1 - {CCF00}\nocite{cote00}; 
2 - \citet{stil02b};
3 - \cite*{meurer98}; 
4 - \cite*{vanzee98c}; 
5 - \cite*{appleton81};
6 - {Carozzi, Chamaraux, \& Duflot-Augarde (1974)}\nocite{carozzi74};
7 - \citet{swaters99};
8 - \cite*{bajaja94}; 
9 - \citet{kobulnicky95};
10 - \citet{schneider92}; 
11 - \citet{sahu97}; 
12 - \cite*{karachentsev99}; 
13 - \citet{devaucouleurs91}; 
14 - \citet{falco99}; 
15 - This work; 
16 - \citet{israel88}; 
17 - \citet{tosi01}; 
18 - \citet{izotov02}; 
19 - \citet{sakai99}; 
20 - \citet{mendez02}; 
21 - \citet{drozdovsky02}; 
22 - \citet{karachentsev98b}; 
23 - \citet{saha95}; 
24 - {MMHS}\nocite{marlowe97}; 
25 - {SCM03b}\nocite{skillman03b}; 
26 - \citet{kobulnicky97b}; 
27 - \citet{storchibergmann95}; 
28 - \citet{skillman94}; 
29 - \citet{read02}; 
30 - \citet{izotov99a}; 
31 - \citet{kobulnicky96};
32 - \cite*{skillman89}; 
33 - \citet{kobulnicky97a}; 
34 - \citet{martin98}; 
35 - {Lehnert, Heckman, \& Weaver (1999)}\nocite{lehnert99}; 
36 - \citet{bomans98}; 
37 - \citet{martin02}; 
38 - \citet{hensler98}; 
39 - \citet{moran97}; 
40 - \citet{papaderos94}; 
41 - \citet{roberts00}; 
42 - \citet{summers03};
43 - \citet{martin95}}
\tablenotetext{a}{The number in parentheses after some entries corresponds to 
the reference from which that data point was taken.}
\tablenotetext{b}{The \HI\ mass quoted in {CCF00}\nocite{cote00} was derived 
assuming a distance of 2.5 Mpc; the \HI\ mass here has been scaled for the new 
distance of 3.89 Mpc.}
\tablenotetext{c}{Derived assuming the distance listed here, and applying the 
mass formula, \begin{math}M_{HI} = 1.7{\times}10^{6}{\times}D^2\end{math} 
\msun, where D is the distance to the galaxy in Mpc, given in 
\citet{carozzi74}.}
\tablenotetext{d}{No uncertainty quoted}
\tablenotetext{e}{Derived from apparent magnitudes listed in NED and using the
distances quoted in this table, unless otherwise noted.}
\tablenotetext{f}{All foreground extinction values are taken from 
{SFD98}\nocite{schlegel98}.}
\tablenotetext{g}{Derived using the \halpha\ fluxes in \citet{martin98}, and 
applying the distance in this table and the conversion to SFR from 
\citet{kennicutt94}.}
\tablenotetext{h}{Calculated using the extinction-corrected total \halpha\ 
luminosity for NGC\,1705 found by {MMHS}\nocite{marlowe97} and applying the 
conversion to SFR from \citet{kennicutt94}}.
\tablenotetext{i}{Calculated using the total \halpha\ fluxes found in 
\citet{skillman94} (UGC\,4483) and \cite*{lehnert99} (M\,82), applying the 
conversion to SFR from \citet{kennicutt94}, and using the distances given in 
Row\,3 of this table.}
\tablenotetext{j}{Fluxes drawn from the NASA/IPAC Infrared Science Archive; 
see http://irsa.ipac.caltech.edu.  100 $\mu$m luminosity derived using the 
distance given in this table, and assuming a 30 $\mu$m effective bandwidth for
the 100 $\mu$m filter (i.e., these values will scale with the adopted filter 
width).}
\label{t7}
\end{deluxetable}

\clearpage
\begin{thebibliography}{}
\bibitem[Allende Prieto, Lambert, \& Asplund(2001)]{allendeprieto01} 
Allende Prieto, C., Lambert, D.~L., \& Asplund, M.\ 2001, \apjl, 556, L63 

\bibitem[Aparicio \etal(1996)]{aparicio96} Aparicio, 
A., Gallart, C., Chiosi, C., \& Bertelli, G.\ 1996, \apjl, 469, L97 

\bibitem[Appleton, Davies, \& Stephenson(1981)]{appleton81} Appleton, P.~N., 
Davies, R.~D., \& Stephenson, R.~J.\ 1981, \mnras, 195, 327 

\bibitem[Bajaja, Huchtmeier, \& Klein(1994)]{bajaja94} Bajaja, E., Huchtmeier,
W.~K., \& Klein, U.\ 1994, \aap, 285, 385 

\bibitem[Bellazzini \etal(2001)]{bellazzini01a} Bellazzini, M.,
Ferraro, F.~R., \& Pancino, E.\ 2001, \apj, 556, 635 

\bibitem[Bertelli \etal(1994)]{bertelli94} Bertelli, G., Bressan, A., Chiosi, 
C., Fagotto, F., \& Nasi, E.\ 1994, \aaps, 106, 275 (B94)

\bibitem[Bomans \etal(2003)]{bomans03} Bomans, D.~J., Cannon, 
J.~M., \& Skillman, E.~D.\ 2003, \apj, in preparation 

\bibitem[Bomans \& Grant(1998)]{bomans98} Bomans, D.~J.~\& Grant, M.-B.\ 1998,
Astronomische Nachrichten, 319, 26 

\bibitem[Bureau \& Carignan(2002)]{bureau02} Bureau, M.~\& Carignan, C.\ 2002,
\aj, 123, 1316 

\bibitem[Cair{\' o}s \etal(2001a)]{cairos01a} Cair{\' o}s, L.~M., 
V{\'{\i}}lchez, J.~M., Gonz{\' a}lez P{\' e}rez, J.~N., Iglesias-P{\' 
a}ramo, J., \& Caon, N.\ 2001a, \apjs, 133, 321

\bibitem[Cair{\' o}s \etal(2001b)]{cairos01b} Cair{\' o}s, L.~M., 
Caon, N., V{\'{\i}}lchez, J.~M., Gonz{\' a}lez-P{\' e}rez, J.~N., \& Mu{\~ 
n}oz-Tu{\~ n}{\' o}n, C.\ 2001b, \apjs, 136, 393 

\bibitem[Calzetti(2001)]{calzetti01} Calzetti, D.\ 2001, \pasp, 113, 1449 

\bibitem[Calzetti \etal(1997)]{calzetti97} Calzetti, D., Meurer, G.~R., 
Bohlin, R.~C., Garnett, D.~R., Kinney, A.~L., Leitherer, C., \& 
Storchi-Bergmann, T.\ 1997, \aj, 114, 1834 

\bibitem[Cannon \etal(2002)]{cannon02} Cannon, J.~M., Skillman, E.~D., 
Garnett, D.~R., \& Dufour, R.~J.\ 2002, \apj, 565, 931 

\bibitem[Carozzi \etal(1974)]{carozzi74} 
Carozzi, N., Chamaraux, P., \& Duflot-Augarde, R.\ 1974, \aap, 30, 21 

\bibitem[Conti(1991)]{conti91} Conti, P.~S.\ 1991, \apj, 377, 115 

\bibitem[C{\^ o}t{\' e} \etal(2000)]{cote00} C{\^ o}t{\' e}, S., Carignan, C.,
\& Freeman, K.~C.\ 2000, \aj, 120, 3027 (CCF00)

\bibitem[Crone \etal(2002)]{crone02} Crone, M.~M., Schulte-Ladbeck, R.~E., 
Greggio, L., \& Hopp, U.\ 2002, \apj, 567, 258 

\bibitem[Da Costa \& Armandroff(1990)]{dacosta90} Da Costa, 
G.~S.~\& Armandroff, T.~E.\ 1990, \aj, 100, 162 

\bibitem[de Vaucouleurs \etal(1991)]{devaucouleurs91} de Vaucouleurs, 
G., de Vaucouleurs, A., Corwin, H.~G., Buta, R.~J., Paturel, G., \& Fouque, 
P.\ 1991, Volume 1-3, XII, 2069 pp.~7 figs..~ Springer-Verlag Berlin 
Heidelberg New York

\bibitem[Dohm-Palmer \etal(1997a)]{dohmpalmer97a} Dohm-Palmer, 
R.~C.~\etal\ 1997a, \aj, 114, 2514 

\bibitem[Dohm-Palmer \etal(1997b)]{dohmpalmer97b} Dohm-Palmer, 
R.~C.~\etal\ 1997b, \aj, 114, 2527 

\bibitem[Dohm-Palmer \etal(1998)]{dohmpalmer98} Dohm-Palmer, 
R.~C.~\etal\ 1998, \aj, 116, 1227 

\bibitem[Dohm-Palmer \etal(2002)]{dohmpalmer02a} Dohm-Palmer, R.~C., 
Skillman, E.~D., Mateo, M., Saha, A., Dolphin, A., Tolstoy, E., Gallagher, 
J.~S., \& Cole, A.~A.\ 2002, \aj, 123, 813 

\bibitem[Dolphin(2000)]{dolphin00b} Dolphin, A.~E.\ 2000, \pasp, 112, 1397 

\bibitem[Dolphin(2002)]{dolphin02b} Dolphin, A.~E.\ 2002, \mnras, 332, 91 

\bibitem[Dolphin \etal(2002)]{dolphin02c} Dolphin, A.~E.~\etal\ 
2002, \aj, 123, 3154 

\bibitem[Dolphin \etal(2001)]{dolphin01} Dolphin, A.~E.~\etal\ 
2001, \apj, 550, 554 

\bibitem[Doublier \etal(1997)]{doublier97} Doublier, V., Comte, 
G., Petrosian, A., Surace, C., \& Turatto, M.\ 1997, \aaps, 124, 405 

\bibitem[Doublier, Caulet, \& Comte(1999)]{doublier99} Doublier, 
V., Caulet, A., \& Comte, G.\ 1999, \aaps, 138, 213 

\bibitem[Drozdovsky \etal(2002)]{drozdovsky02} Drozdovsky, I.~O., 
Schulte-Ladbeck, R.~E., Hopp, U., Greggio, L., \& Crone, M.~M.\ 2002, \aj, 
124, 811 

\bibitem[Dutil \& Roy(2001)]{dutil01} Dutil, Y.~\& Roy, J.\ 
2001, \aj, 122, 1644 

\bibitem[Falco \etal(1999)]{falco99} Falco, E.~E.~\etal\ 
1999, \pasp, 111, 438 

\bibitem[Ferguson \& Babul(1998)]{ferguson98} Ferguson, H.~C.~\& 
Babul, A.\ 1998, \mnras, 296, 585 

\bibitem[Ferrarese \etal(2000)]{ferrarese00a} Ferrarese, L.~et al.\ 
2000, \apj, 529, 745 

\bibitem[Gallagher \etal(1998)]{gallagher98} Gallagher, J.~S., 
Tolstoy, E., Dohm-Palmer, R.~C., Skillman, E.~D., Cole, A.~A., Hoessel, 
J.~G., Saha, A., \& Mateo, M.\ 1998, \aj, 115, 1869 

\bibitem[Gallart \etal(1994)]{gallart94} Gallart, C., Aparicio, 
A., Chiosi, C., Bertelli, G., \& Vilchez, J.~M.\ 1994, \apjl, 425, L9 

\bibitem[Gallart, Aparicio, \& Vilchez(1996)]{gallart96a} Gallart, 
C., Aparicio, A., \& Vilchez, J.~M.\ 1996, \aj, 112, 1928 

\bibitem[Garnett(1990)]{garnett90} Garnett, D.~R.\ 1990, \apj, 
363, 142 

\bibitem[Gil de Paz \etal(2003)]{gildepaz03} Gil de Paz, A.,
Madore, B.~F., \& Pevunova, O.\ 2003, \apjs, 147, 29

\bibitem[Greggio \etal(1998)]{greggio98} Greggio, L., Tosi, M., 
Clampin, M., de Marchi, G., Leitherer, C., Nota, A., \& Sirianni, M.\ 1998, 
\apj, 504, 725 

\bibitem[Gonz{\' a}lez Delgado, Leitherer, \& 
Heckman(1999)]{gonzalezdelgado99b} Gonz{\' a}lez Delgado, R.~M., 
Leitherer, C., \& Heckman, T.~M.\ 1999, \apjs, 125, 489 

\bibitem[Heckman(1998)]{heckman98conf} Heckman, T.~M.\ 1998, ASP 
Conf.~Ser.~148: Origins, 127 

\bibitem[Hensler \etal(1998)]{hensler98} Hensler, G., Dickow, R., Junkes, 
N., \& Gallagher, J.~S.\ 1998, \apjl, 502, L17 

\bibitem[Hensler \& K{\" o}ppen(1999)]{hensler99} Hensler, G.~\& 
K{\" o}ppen, J.\ 1999, Astronomische Gesellschaft Meeting Abstracts, 15, 16 

\bibitem[Hodge(1989)]{hodge89} Hodge, P.\ 1989, \araa, 27, 139 

\bibitem[Holtzman \etal(1995)]{holtzman95b} Holtzman, J.~A., 
Burrows, C.~J., Casertano, S., Hester, J.~J., Trauger, J.~T., Watson, 
A.~M., \& Worthey, G.\ 1995, \pasp, 107, 1065 

\bibitem[Hopkins, Schulte-Ladbeck, \& Drozdovsky(2002)]{hopkins02} 
Hopkins, A.~M., Schulte-Ladbeck, R.~E., \& Drozdovsky, I.~O.\ 2002, \aj, 
124, 862 

\bibitem[Hummer \& Storey(1987)]{hummer87} Hummer, D.~G.~\& 
Storey, P.~J.\ 1987, \mnras, 224, 801 

\bibitem[Hunter \etal(2000)]{hunter00} Hunter, D.~A., O'Connell, R.~W., 
Gallagher, J.~S., \& Smecker-Hane, T.~A.\ 2000, \aj, 120, 2383 

\bibitem[Israel(1988)]{israel88} Israel, F.~P.\ 1988, \aap, 194, 24 

\bibitem[Izotov \& Thuan(1999)]{izotov99a} Izotov, Y.~I.~\& 
Thuan, T.~X.\ 1999, \apj, 511, 639 

\bibitem[Izotov \& Thuan(2002)]{izotov02} Izotov, Y.~I.~\& 
Thuan, T.~X.\ 2002, \apj, 567, 875 

\bibitem[Karachentsev \& Drozdovsky(1998)]{karachentsev98b} 
Karachentsev, I.~D.~\& Drozdovsky, I.~O.\ 1998, \aaps, 131, 1 

\bibitem[Karachentsev, Makarov, \& Huchtmeier(1999)]{karachentsev99} 
Karachentsev, I.~D., Makarov, D.~I., \& Huchtmeier, W.~K.\ 1999, \aaps, 
139, 97 

\bibitem[Karachentsev \etal(2003)]{karachentsev03b} Karachentsev, 
I.~D.~et al.\ 2003, \aap, 404, 93 

\bibitem[Kennicutt \& Skillman(2001)]{kennicutt01} Kennicutt, 
R.~C.~\& Skillman, E.~D.\ 2001, \aj, 121, 1461 

\bibitem[Kennicutt \etal(1994)]{kennicutt94} Kennicutt, R.~C., 
Tamblyn, P., \& Congdon, C.~E.\ 1994, \apj, 435, 22 

\bibitem[Kobulnicky \& Skillman(1995)]{kobulnicky95} Kobulnicky, 
H.~A.~\& Skillman, E.~D.\ 1995, \apjl, 454, L121 

\bibitem[Kobulnicky \& Skillman(1996)]{kobulnicky96} Kobulnicky, 
H.~A.~\& Skillman, E.~D.\ 1996, \apj, 471, 211 

\bibitem[Kobulnicky \& Skillman(1997)]{kobulnicky97b} Kobulnicky, 
H.~A.~\& Skillman, E.~D.\ 1997, \apj, 489, 636 

\bibitem[Kobulnicky \etal(1997)]{kobulnicky97a} Kobulnicky, H.~A., Skillman, 
E.~D., Roy, J., Walsh, J.~R., \& Rosa, M.~R.\ 1997, \apj, 477, 679 

\bibitem[Krueger, Fritze-v.~Alvensleben, \& Loose(1995)]{krueger95} Krueger, 
H., Fritze-v.~Alvensleben, U., \& Loose, H.-H.\ 1995, \aap, 303, 41 

\bibitem[Kunth \& Sargent(1981)]{kunth81} Kunth, D.~\& Sargent, 
W.~L.~W.\ 1981, \aap, 101, L5 

\bibitem[Lee \etal(1993)]{lee93} Lee, M.~G., 
Freedman, W.~L., \& Madore, B.~F.\ 1993, \apj, 417, 553 

\bibitem[Lehnert \etal(1999)]{lehnert99} Lehnert, 
M.~D., Heckman, T.~M., \& Weaver, K.~A.\ 1999, \apj, 523, 575 

\bibitem[Loose \& Thuan(1986)]{loose86} Loose, H.~H.~\& Thuan, 
T.~X.\ 1986, Star Forming Dwarf Galaxies and Related Objects, 73 

\bibitem[Lynds \etal(1998)]{lynds98} Lynds, R., Tolstoy, E., O'Neil., E.~J., 
\& Hunter, D.~A.\ 1998, \aj, 116, 146 

\bibitem[Madore \& Freedman(1995)]{madore95} Madore, B.~F.~\& 
Freedman, W.~L.\ 1995, \aj, 109, 1645 

\bibitem[Marlowe \etal(1997)]{marlowe97} Marlowe, A.~T., Meurer, G.~R., 
Heckman, T.~M., \& Schommer, R.\ 1997, \apjs, 112, 285 (MMHS)

\bibitem[Marlowe, Meurer, \& Heckman(1999)]{marlowe99} Marlowe, 
A.~T., Meurer, G.~R., \& Heckman, T.~M.\ 1999, \apj, 522, 183 

\bibitem[Martin \& Kennicutt(1995)]{martin95} Martin, C.~L.~\& 
Kennicutt, R.~C.\ 1995, \apj, 447, 171 

\bibitem[Martin(1998)]{martin98} Martin, C.~L.\ 1998, \apj, 506, 222 

\bibitem[Martin \etal(2002)]{martin02} Martin, 
C.~L., Kobulnicky, H.~A., \& Heckman, T.~M.\ 2002, \apj, 574, 663 

\bibitem[Mateo(1998)]{mateo98} Mateo, M.~L.\ 1998, \araa, 36, 435 

\bibitem[M{\' e}ndez \etal(2002)]{mendez02} M{\' e}ndez, B., 
Davis, M., Moustakas, J., Newman, J., Madore, B.~F., \& Freedman, W.~L.\ 
2002, \aj, 124, 213 

\bibitem[Meurer, Staveley-Smith, \& Killeen(1998)]{meurer98} Meurer, G.~R., 
Staveley-Smith, L., \& Killeen, N.~E.~B.\ 1998, \mnras, 300, 705 

\bibitem[Miller(1996)]{miller96} Miller, B.~W.\ 1996, \aj, 112, 991 

\bibitem[Minniti \& Zijlstra(1997)]{minniti97} Minniti, D.~\& 
Zijlstra, A.~A.\ 1997, \aj, 114, 147 

\bibitem[Moran \& Lehnert(1997)]{moran97} Moran, E.~C.~\& 
Lehnert, M.~D.\ 1997, \apj, 478, 172 

\bibitem[Papaderos \etal(1994)]{papaderos94} Papaderos, P., Fricke, K.~J., 
Thuan, T.~X., \& Loose, H.-H.\ 1994, \aap, 291, L13 

\bibitem[Papaderos \etal(1996a)]{papaderos96a} 
Papaderos, P., Loose, H.-H., Thuan, T.~X., \& Fricke, K.~J.\ 1996a, \aaps, 
120, 207 

\bibitem[Papaderos \etal(1996b)]{papaderos96b} 
Papaderos, P., Loose, H.-H., Fricke, K.~J., \& Thuan, T.~X.\ 1996b, \aap, 
314, 59 

\bibitem[Puche \& Carignan(1988)]{puche88} Puche, D.~\& 
Carignan, C.\ 1988, \aj, 95, 1025 

\bibitem[Putman \etal(2002)]{putman02} Putman, M.~E.~\etal\ 
2002, \aj, 123, 873 

\bibitem[Ratnatunga \& Bahcall(1985)]{ratnatunga85} Ratnatunga, 
K.~U.~\& Bahcall, J.~N.\ 1985, \apjs, 59, 63 

\bibitem[Read \& Stevens(2002)]{read02} Read, A.~M.~\& 
Stevens, I.~R.\ 2002, \mnras, 335, L36 

\bibitem[Roberts \& Warwick(2000)]{roberts00} Roberts, T.~P.~\& 
Warwick, R.~S.\ 2000, \mnras, 315, 98 

\bibitem[Romaniello \etal(2002)]{romaniello02} Romaniello, M., 
Panagia, N., Scuderi, S., \& Kirshner, R.~P.\ 2002, \aj, 123, 915 

\bibitem[Saha \etal(1995)]{saha95} Saha, A., Sandage, A., Labhardt, L., 
Schwengeler, H., Tammann, G.~A., Panagia, N., \& Macchetto, 
F.~D.\ 1995, \apj, 438, 8 

\bibitem[Saha \etal(1996)]{saha96} Saha, A., Sandage, A., Labhardt, L., 
Tammann, G.~A., Macchetto, F.~D., \& Panagia, N.\ 1996, \apj, 466, 55 

\bibitem[Sahu \& Blades(1997)]{sahu97} Sahu, M.~S.~\& Blades, 
J.~C.\ 1997, \apjl, 484, L125 

\bibitem[Sakai \& Madore(1999)]{sakai99} Sakai, S.~\& Madore, 
B.~F.\ 1999, \apj, 526, 599 

\bibitem[Sakai \etal(1996)]{sakai96} Sakai, S., 
Madore, B.~F., \& Freedman, W.~L.\ 1996, \apj, 461, 713 

\bibitem[Schaerer \etal(1999a)]{schaerer99a} Schaerer, 
D., Contini, T., \& Kunth, D.\ 1999a, \aap, 341, 399 

\bibitem[Schaerer \etal(1999b)]{schaerer99b} Schaerer, 
D., Contini, T., \& Pindao, M.\ 1999b, \aaps, 136, 35 

\bibitem[Schaerer \& Vacca(1998)]{schaerer98} Schaerer, D.~\& 
Vacca, W.~D.\ 1998, \apj, 497, 618 

\bibitem[Schechter, Mateo, \& Saha(1993)]{schechter93} Schechter, 
P.~L., Mateo, M., \& Saha, A.\ 1993, \pasp, 105, 1342 

\bibitem[Schlegel \etal(1998)]{schlegel98} Schlegel, D.~J., 
Finkbeiner, D.~P., \& Davis, M.\ 1998, \apj, 500, 525 (SFD98)

\bibitem[Schneider \etal(1992)]{schneider92} Schneider, S.~E., Thuan, T.~X., 
Mangum, J.~G., \& Miller, J.\ 1992, \apjs, 81, 5 

\bibitem[Schulte-Ladbeck, Crone, \& Hopp(1998)]{schulteladbeck98} 
Schulte-Ladbeck, R.~E., Crone, M.~M., \& Hopp, U.\ 1998, \apjl, 493, L23 

\bibitem[Schulte-Ladbeck \etal(1999)]{schulteladbeck99a} Schulte-Ladbeck, 
R.~E., Hopp, U., Crone, M.~M., \& Greggio, L.\ 1999, \apj, 525, 709 

\bibitem[Searle \& Sargent(1972)]{searle72} Searle, L.~\& 
Sargent, W.~L.~W.\ 1972, \apj, 173, 25 

\bibitem[Searle, Sargent, \& Bagnuolo(1973)]{searle73} Searle, 
L., Sargent, W.~L.~W., \& Bagnuolo, W.~G.\ 1973, \apj, 179, 427 

\bibitem[Skillman, Bomans, \& Kobulnicky(1997)]{skillman97} Skillman, E.~D., 
Bomans, D.~J., \& Kobulnicky, H.~A.\ 1997, \apj, 474, 205 

\bibitem[Skillman \etal(2003a)]{skillman03a} Skillman, E.~D., C{\^ o}t{\' e}, 
S., \& Miller, B.~W.\ 2003a, \aj, 125, 593 (SCM03a)

\bibitem[Skillman \etal(2003b)]{skillman03b} Skillman, E.~D., C{\^ o}t{\' e}, 
S., \& Miller, B.~W.\ 2003b, \aj, 125, 610 (SCM03b)

\bibitem[Skillman, Kennicutt, \& Hodge(1989)]{skillman89} Skillman, E.~D., 
Kennicutt, R.~C., \& Hodge, P.~W.\ 1989, \apj, 347, 875 

\bibitem[Skillman \etal(1994)]{skillman94} Skillman, E.~D., Terlevich, R.~J., 
Kennicutt, R.~C., Garnett, D.~R., \& Terlevich, E.\ 1994, \apj, 431, 172 

\bibitem[Stil \& Israel(2002)]{stil02b} Stil, J.~M.~\& Israel, 
F.~P.\ 2002, \aap, 392, 473 

\bibitem[Storchi-Bergmann, Kinney, \& Challis(1995)]{storchibergmann95} 
Storchi-Bergmann, T., Kinney, A.~L., \& Challis, P.\ 1995, \apjs, 98, 103 

\bibitem[Summers \etal(2003)]{summers03} Summers, L.~K., Stevens, I.~R., 
Strickland, D.~K. and Heckman, T.~M.\ 2003, \mnras, 342, 690

\bibitem[Swaters(1999)]{swaters99} Swaters, R.~A.\ 1999, Ph.D.~Thesis, 
Rijksuniversiteit Groningen

\bibitem[Taylor(1997)]{taylor97} Taylor, C.~L.\ 1997, \apj, 480, 
524 

\bibitem[Thuan \& Martin(1981)]{thuan81} Thuan, T.~X.~\& 
Martin, G.~E.\ 1981, \apj, 247, 823 

\bibitem[Tolstoy \etal(1998)]{tolstoy98} Tolstoy, E.~et al.\ 
1998, \aj, 116, 1244 

\bibitem[Tosi \etal(2001)]{tosi01} Tosi, M., Sabbi, E., Bellazzini, M., 
Aloisi, A., Greggio, L., Leitherer, C., \& 
Montegriffo, P.\ 2001, \aj, 122, 1271 

\bibitem[Tremonti \etal(2001)]{tremonti01} Tremonti, C.~A., Calzetti, D., 
Leitherer, C., \& Heckman, T.~M.\ 2001, \apj, 555, 322 

\bibitem[van Zee(2000)]{vanzee00a} van Zee, L.\ 2000, \aj, 119, 2757 

\bibitem[van Zee, Skillman, \& Salzer(1998)]{vanzee98c} van Zee, 
L., Skillman, E.~D., \& Salzer, J.~J.\ 1998, \aj, 116, 1186 

\bibitem[Whitmore \etal(1999)]{whitmore99} 
Whitmore, B., Heyer, I., \& Casertano, S.\ 1999, \pasp, 111, 1559 

\bibitem[Zaritsky(1999)]{zaritsky99} Zaritsky, D.\ 1999, \aj, 118, 2824 

\bibitem[Zaritsky \etal(2002)]{zaritsky02} Zaritsky, D., Harris, 
J., Thompson, I.~B., Grebel, E.~K., \& Massey, P.\ 2002, \aj, 123, 855 
\end{thebibliography}
\end{document}